\documentclass[letter,11pt,round]{article}

\RequirePackage[OT1]{fontenc}
\usepackage{amsthm,amsmath,natbib}
\RequirePackage{hypernat}
\usepackage[pdftex]{graphicx}
\usepackage{amsfonts}
\usepackage{epic}
\usepackage{fancybox}
\usepackage[pdftex]{color}
\usepackage{marginnote}
\usepackage{multirow}
\usepackage{soul}
\usepackage{adjustbox}

\usepackage[letterpaper]{geometry}
\RequirePackage{ifthen}

\usepackage{amsmath}
\DeclareMathOperator\arctanh{arctanh}
\DeclareMathOperator\atanh{arctanh}

\numberwithin{equation}{section}
\theoremstyle{plain}
\newtheorem{thm}{Theorem} 
\newtheorem{lemma}{Lemma} 
\newtheorem{prop}{Proposition} 
\newtheorem{remark}{Remark}[section]
\newtheorem{definition}{Definition}[section]

\definecolor{gray}{rgb}{0.3,0.3,0.3}

\ifthenelse{1=1}{}{
\usepackage[backend=biber]{biblatex}

\NewBibliographyString{diplomathesis}
\DefineBibliographyStrings{english}{
  diplomathesis = {diploma thesis},
}

\addbibresource{\jobname.bib}

\nocite{*}
}

\begin{document}

\bibliographystyle{chicago}

\newcommand{\replace}[2]{#2}

\title{Reweighted and Circularized Anderson-Darling Tests of Goodness-of-Fit}

\author{
  Chuanhai Liu\\
	\\
	Department of Statistics, Purdue University
	\\ E-mail: chuanhai@purdue.edu
}
	\date{\today\thanks{To appear in Journal of Nonparametric Statistics.}}

\maketitle

\newcommand{\prsbound}{B}

\newcommand{\Gdist}{\mathbb{G}}

\begin{abstract}
This paper takes a look at omnibus tests of goodness of fit in the context of reweighted Anderson-Darling tests and makes threefold contributions.  The first contribution is to provide a geometric understanding. It is argued that the test statistic with minimum variance for exchangeable distributional deviations can serve as a good general-purpose test.  The second contribution is to propose better omnibus tests, called circularly symmetric tests and obtained by circularizing reweighted Anderson-Darling test statistics or, more generally, test statistics based on the observed order statistics. The resulting tests are called circularized tests. A limited but arguably convincing simulation study on finite-sample performance demonstrates that circularized tests have good performance, as they typically outperform their parent methods in the simulation study. The third contribution is to establish new large-sample results.

	\vspace{0.2in}


\end{abstract}

\section{Introduction}\label{s:intro}

The problem of determining whether a sample of $n$ observations
$X_1$,...,$X_n$ can be considered as a sample from a 
given continuous distribution $F(x)$, 
known as goodness of fit (GOF), is theoretically fundamental.
It is also practically important, especially
for contemporary big-data analysis, for
model building and model checking in particular and nonparametric inference in general.
The methodological development for assessing
goodness of fit has been a significant part of statistical research 
in the past century.  It can be traced
back to Pearson's chi-square test \citep{pearson1900x}
and has made available many influential methods,
including the Cram\'er-von Mises criterion \citep{cramer1928composition,vonmises1928Wahr},
Kolmogorov-Smirnov test \citep{kolmogorov1933sulla,smirnov1939deviation},
Anderson-Darling test \citep{anderson1952asymptotic,anderson1954test},
Shapiro-Wilk test \citep{shapiro1965analysis},
and Zhang's likelihood-ratio test \citep{zhang2002powerful}; see
Lehmann and Romano (2005, pp. 629-630)
for a comprehensive list of references.
Among these classical tests, Anderson-Darling and Zhang have been perceived 
as powerful \cite[see, {\it e.g.},][]{sinclair1988approximations,zhang2010statistical}.

The Anderson-Darling test \citep{anderson1952asymptotic} is defined as a weighted
empirical distribution statistic
\begin{equation}\label{eq:EDF-01}
A_n^2(w) = n\int_{-\infty}^{\infty}\left[F_n(x) - F(x)\right]^2w(x)dx
\end{equation}
with the null distribution $F(.)$ and 
the weight function $w(x)= \frac{1}{F(x)(1-F(x))}$, where
$F_n(.)$ denotes the usual empirical distribution function:
\[ F_n(x) = \frac{k}{n}\quad \mbox{ if $k$ observations
are $\le x$}.
	\]
In comparing Anderson-Darling and Cram\'er-von Mises, 
\cite{anderson1952asymptotic} wrote:
\begin{quote}
	\it
	A statistician may prefer to use this weight function [$\psi(x)
	=1/[F(x)(1-F(x))]$] 
 when he feels that $\psi(x)=1$
	does not give enough weight to the tails of the distribution
	[$F(x)$].
\end{quote}
Although this is true, the comparison is still only relative.
The most important to an applied statistician
is perhaps that the question
`{\it what would be a default or all-purpose test that could be
considered relatively neutral regarding the location of deviations from the hypothesized distribution?}' remains, however, to be answered.
This paper takes a look at this so-called omnibus testing problem
and aims at three goals.

Let $F^*(.)$ denote the underlying true distribution function of the observed sample $X_1,..., X_n$.
The first goal of this paper is to develop geometric intuitions
and the corresponding mathematical theory toward an answer
to the above question
by considering a class of reweighted Anderson-Darling tests.
We establish an explicit
one-to-one correspondence between the weights and their focal directions of the distributional deviations of $F^*(.)$ from $F(.)$ at the sorted values of $X_1,..., X_n$.
For example, it is found that the weights that produce the test statistic with minimal variance assign equal importance to all standardized deviations.
As a result, we take the corresponding test as a general-purpose test.
This arguably optimal weight-based test is found to be 
similar to the Zhang test, which has been widely perceived as powerful.
Our geometric arguments and the corresponding theoretical results in Section \ref{s:DA-review} offer an additional perspective on understanding its performance.

The second goal is to explore better omnibus tests.
It is recognized that
existing powerful methods suffer from the confounded effect of
locations and frequencies in the deviations from the null hypothesis.
This motivates the proposed circularization method
to create circularly symmetric tests by
circularizing reweighted Anderson-Darling test statistics and, more generally, any test statistic based on the order statistics of the observed sample $X_1,..., X_n$.
Two types of circularization are considered: one
is obtained by taking the average of the corresponding
statistics and the other by using the maximum.
A simple but arguably convincing simulation study
in Section \ref{s:power-comparison}
on finite-sample performance demonstrates
that the circularized Zhang method outperforms the
circularized Anderson-Darling method and that the circularized tests
outperform their parent methods.

The final goal is to establish new large-sample results.
It is found that, like Anderson-Darling, the test statistics under the null hypothesis
have the same distribution as that of a weighted sum of
an infinite number of independent squared normal random variables.
These theoretical results are shown numerically to be useful for
large sample-based approximations.
It should be noted that our exploration focuses mostly on statistical ideas and geometric intuitions. For this reason, most of our arguments are made using approximations, except those stated formally in theorems.

The rest of the paper is arranged as follows.
Section \ref{s:DA-review} introduces the basic notation and
the class of reweighted Anderson-Darling test statistics.
Section \ref{s:optim-weights} develops
statistical intuitions for understanding
reweighted Anderson-Darling tests.  The default or optimal weights are then defined
accordingly, followed by
an investigation of finite-sample cases
and a large-sample theory-based solution.
Section \ref{s:cs} discusses the difficulties suffered by the
existing powerful methods and proposes
circularly symmetric tests.  Section \ref{s:power-comparison} 
considers a simple simulation study which demonstrates
that the circularized Zhang method outperforms the
circularized Anderson-Darling method and that the circularized tests
outperform their parent methods.
Section \ref{s:asym-dist} discusses
the limiting distributions of three proposed test statistics.
Section \ref{s:conclusion} concludes
with a few remarks.

\section{Reweighted Anderson-Darling Tests}
\label{s:DA-review}

\subsection{Reweighted Anderson-Darling tests}
The basic setting for the theoretical investigation is
that independent and identically distributed
random variables $X_1$,...,$X_n$ have a specified continuous distribution
$F(x)$, $x\in \mathbb{R}$.
Denote by $X_{(1)} \leq  ... \le  X_{(n)}$ the corresponding order statistics.
This set of order statistics or the corresponding
order statistics $U_{(1)}=F(X_{(1)}) \leq  ... \le  U_{(n)}=F(X_{(n)})$
are sufficient for inference about $F(.)$,
especially when inference about unknown $F(.)$ is of interest.
It is well known and easy to prove that 
under the null, 
the sampling distribution of $U_{(1)}, ...,  U_{(n)}$
is that of a sorted uniform sample of size $n$. 
In the context of hypothesis testing,
we write the null hypothesis as
\[H_0: F^*(x)=F(x)\quad \mbox{ for all } x\in\mathbb{R}
\]
and the alternative as
\[H_1: F^*(x)\neq F(x)
\quad \mbox{ for some } x\in\mathbb{R}
\] 
where
$F^*(.)$ stands for the true distribution of the observed sample
$X_1$,...,$X_n$.  

Let 
$a_i = \frac{i-\frac{1}{2}}{n}$ for
$i=1,...,n$.  The Anderson-Darling statistic can be written simply and equivalently as
\begin{equation}\label{eq:AD-00}
	\replace{\bar{A}_n^2 = - 2\sum_{i=1}^n\left[ a_i\ln\frac{U_{(i)}}{a_i}
		+(1-a_i)\ln\frac{1-U_{(i)}}{1-a_i}
		\right],}{
  A_n^2 = - 2\sum_{i=1}^nw_i\left[ a_i^2(1-a_i)\ln\frac{U_{(i)}}{a_i}
		+a_i(1-a_i)^2\ln\frac{1-U_{(i)}}{1-a_i}
		\right],
        }
\end{equation}
\replace{making it statistically more intuitive in terms of the finite number of sufficient statistics $U_{(1)},...,U_{(n)}$.}{where $w_i = \frac{1}{a_i(1-a_i)}$ for $i=1,..., n.$}
Note that under the null $H_0$, $\mu_i = E(U_{(i)})= \frac{i}{n+1}$, for $i=1,...,n.$
For theoretical convenience in using moments for developing geometric interpretation,
we replace $a_i$ 
in \eqref{eq:AD-00} with
$\mu_i$ and consider the slightly modified version:
\begin{equation}\label{eq:AD-01}
	W_n^2 = - 2
	\sum_{i=1}^n\replace{\left[
		\mu_i\ln\frac{U_{(i)}}{\mu_i}
		+(1-\mu_i)\ln\frac{1-U_{(i)}}{1-\mu_i}
		\right].}{w_i
\left[
		\mu_i^2(1-\mu_i)\ln\frac{U_{(i)}}{\mu_i}
		+\mu_i(1-\mu_i)^2\ln\frac{1-U_{(i)}}{1-\mu_i}
		\right],}
\end{equation}
\replace{}{where $w_i = \frac{1}{\mu_i(1-\mu_i)}$ for $i=1,..., n.$}
That is, this is done in the analogy with methods using the alternative definition of
empirical distribution
\[F_n(x) = \frac{\left|\left\{i: X_i \leq x\right\}\right|}{n+1},
\qquad(x\in (-\infty, \infty)).\]
where $\left|\left\{i: X_i \leq x\right\}\right|$ is the
number of $X_i$s that are less than or equal to $x$.

The modified version \eqref{eq:AD-01} has 
a very simple and intuitive interpretation.
Note that the marginal probability density function (pdf)
of $U_{(i)}$ is $\mbox{Beta}(i, n+1-i)$, the Beta
distribution with the two shape parameters $i$ and $n+1-i$ \citep[see, {\it e.g.},][p.14]{david2004order}.
Let $Z_i = \ln\frac{U_{(i)}}{1-U_{(i)}}$, the logit transformation
of $U_{(i)}$.
The $i$-th summand of $W_n^2$ is proportional to the negative log probability density function (pdf).
This implies that the Anderson-Darling test statistic
is approximately the average of squares of standardized $Z_i$'s,
which is stochastically small under the null hypothesis and 
large under the alternative.
This can be seen more easily with the following
approximation to the $i$-th summand of $W_n^2$
via the second-order Taylor expansion in terms of
$U_{(i)}$ at $U_{(i)}=\mu_i$:
\begin{equation}\label{eq:AD-02}
	\replace{Y_i \equiv
	-2\left[
\replace{\mu_i}{\mu^2(1-\mu_i)}\ln\frac{U_{(i)}}{\mu_i} +\replace{(1-\mu_i)}{\mu_i(1-\mu_i)^2}\ln\frac{1-U_{(i)}}{1-\mu_i}
	\right]
	\approx \frac{1}{(n+2)}
	\frac{\left(U_{(i)}-\mu_i\right)^2}{\mbox{Var}\left(U_{(i)}\right)}}
 {
 Y_i \equiv
	-2\left[
\mu_i^2(1-\mu_i)\ln\frac{U_{(i)}}{\mu_i} +\mu_i(1-\mu_i)^2\ln\frac{1-U_{(i)}}{1-\mu_i}
	\right]
	\approx \left(U_{(i)}-\mu_i\right)^2,
 }
\end{equation}
where $\mbox{Var}\left(U_{(i)}\right) = \frac{\mu_i(1-\mu_i)}{n+2}$,
under the null and going to zero as $n\rightarrow \infty$.

Recall that 
Cram\'er-von Mises test statistic $\omega^2_n$ is given by \citep{anderson1952asymptotic}
\[
n\omega^2_n =n\int_{-\infty}^\infty [F_n(x)-F(x)]^2dF
= \frac{1}{12n}+\sum_{i=1}^n\left(U_{(i)}-a_i\right)^2.
\]
\ifthenelse{1=1}{}{
which can be obtained from \citep{anderson1952asymptotic}
\begin{eqnarray*}
&&n\int_{-\infty}^\infty [F_n(x)-F(x)]^2dF\\
&=&n\int_0^1[G_n(u)-u]^2du\\
&=& n\sum_{i=0}^n\int_{U_{(i)}}^{U_{(i+1)}}[i/n-u]^2du\\
&=& n\sum_{i=0}^n \left[(i/n)^2(U_{(i+1)}-U_{(i)})-i/n[u_{i+1}^2-u_i^2]\right]+n\int_0^1u^2du\\
&=& n - \sum_{i=0}^n 2a_iu_i+\sum_{i=1}^nu_i^2-n+n/3\\
&=& n/3 - \sum_{i=0}^n a_i^2+\sum_{i=1}^n(u_i^2-a_i)^2\\
&=& n/3 - \frac{1}{n^2}\left[n(n+1)(2n+1)/6-n(n+1)/2+n/4\right]+\sum_{i=1}^n(u_i^2-a_i)^2\\
&=& n/3 - \frac{1}{n}\left[(n+1)(2n+1)/6-(n+1)/2+1/4\right]+\sum_{i=1}^n(u_i^2-a_i)^2\\
&=& \frac{n}{12}+\sum_{i=1}^n(u_i^2-a_i)^2\\
\end{eqnarray*}
\clearpage
where $U_{(0)}=0$, $U_{(n+1)}=1$,
and $G_n(u)$ denotes the empirical distribution of the sample $F(X_i)$, $i=1,...,n.$}
\noindent
Let 
\[D_n = 
\frac{1}{12n}+\sum_{i=1}^n\left(U_{(i)}-a_i\right)^2-\left[\frac{1}{12n}+\sum_{i=1}^n\left(U_{(i)}-\mu_i\right)^2\right]
\]
It can be shown that under $H_0$, both the mean and the variance of $D_n$ converge to zero as $n$ goes to infinity.
Applying Chebyshev's inequality leads to the fact that $D_n$ converges to zero in probability.
It is known \citep[see, {\it e.g.},][]{anderson1952asymptotic} that $n\omega^2$ converges in distribution. This implies that Cram\'er-von Mises criterion
is asymptotically 
the straight average of the squared deviations
$\left(U_{(i)}-\mu_i\right)^2$:
\[\frac{1}{12n}+\sum_{i=1}^n\left(U_{(i)}-\mu_i\right)^2
= n\omega^2  - D_n
\]
because, by Slutsky's theorem, $n\omega^2  - D_n$ converges in distribution to the same limiting distribution of $n\omega^2$.
Thus, compared to Cram\'er-von Mises test statistic,
Anderson-Darling is the weighted average of
the squared deviations $\left(U_{(i)}-\mu_i\right)^2$
with the weights inversely proportional to the variance
of the deviations $U_{(i)}-\mu_i$; see \eqref{eq:AD-01} and \eqref{eq:AD-02}. 

Notice that the deviations
$U_{(i)}-\mu_i$ and, thereby, their squared versions
$\left(U_{(i)}-\mu_i\right)^2$ near the central area of $F(x)$ 
are more correlated than those in the tails.
It is worth considering to weight the tail areas even more 
than Anderson-Darling.
This motivates us to consider the following class of
reweighted Anderson-Darling test statistics:
\begin{equation}\label{eq:generalized-AD-01}
	R_n^2(w) = - 2 
	\sum_{i=1}^n w_i \replace{\left[
		\mu_i\ln \frac{ U_{(i)}}{\mu_i}+(1-\mu_i)\ln\frac{1-U_{(i)}}{1-\mu_i}\right]
  }{
  \left[
		\mu_i^2(1-\mu_i)\ln \frac{ U_{(i)}}{\mu_i}+\mu_i(1-\mu_i)^2\ln\frac{1-U_{(i)}}{1-\mu_i}\right]
  =\sum_{i=1}^{n} w_iY_i
  }
\end{equation}
where $w_i \ge 0$ and $Y_i$ is defined in \eqref{eq:AD-02} for $i=1,...,n$.

The special case of the test statistic \eqref{eq:generalized-AD-01} with weights $w_i=1$ for $i=1,...,n$ corresponds to the Cram\'er-von Mises test statistic, while the case with $w_i=1/[\mu_i(1-\mu_i)]$ corresponds to the 
slightly modified Anderson-Darling test statistic \eqref{eq:AD-01}. 
The default or optimal weights, so-called in this paper and defined for terminology convenience
and studied in detail in
Section \ref{s:optim-weights}, are found to be
\begin{equation}\label{eq:opti-weight-00}
	w_i\propto \replace{\frac{1}{\mu_i(1-\mu_i)}}{
\frac{1}{\mu_i^2(1-\mu_i)^2}
 }
	\qquad (i=1,...,n)
\end{equation}
for large $n$ and weight tails slightly more for small $n$.
This leads to the following test statistic
\begin{equation}\label{eq:optim-test-01}
	R_n^2 = - 2C_n 
	\sum_{i=1}^n \replace{\frac{1}{\mu_i(1-\mu_i)} \left[
		\mu_i\ln \frac{ U_{(i)}}{\mu_i}+(1-\mu_i)\ln\frac{1-U_{(i)}}{1-\mu_i}\right]}
  {
\left[
		\frac{1}{1-\mu_i}\ln \frac{ U_{(i)}}{\mu_i}+\frac{1}{\mu_i}\ln\frac{1-U_{(i)}}{1-\mu_i}\right]
  }
\end{equation}
where the rescaling constant $C_n=\frac{1}{2\sum_{k=1}^n\frac{1}{k}}$
is taken for the preference of $E(R_n^2)\approx 1$
(see Section \ref{s:limiting-dist}).

Clearly,  
this test is similar to the likelihood ratio (LR) test statistic of \cite{zhang2002powerful},
as they are equivalent when $\mu_i=i/(n+1)$ in \eqref{eq:optim-test-01} is replaced by $a_i=(i-1/2)/n$
\begin{equation}\label{eq:zhang-test-01}
Z_A = -\sum_{i=1}^n \left[
	\frac{\ln ( U_{(i)})}{n-i+\frac{1}{2}}
	+ \frac{\ln (1- U_{(i)})}{i-\frac{1}{2}}\right],
\end{equation}
which is obtained by weighting the likelihood ratios
for the individual $F(t)$ instead of $[F_n(x)-F(x)]^2$
in \eqref{eq:EDF-01} with the choice of the weight function
$1/[F(x)(1-F(x))]$. 
It should be noted that the Zhang test has been
commonly perceived as powerful
\cite[see, {\it e.g},][]{zhang2002powerful,zhang2010statistical,carmen2022extensions}. 
The good performance of $R_n^2$ compared to $W_n^2$
is also demonstrated 
in Section \ref{s:power-comparison}.
This helps to see the performance of $R_n^2$, in addition to
the geometric interpretation provided in Section \ref{s:optim-weights}.
In addition, 
the geometric intuitions and the corresponding theoretical
investigation in the next section shed light on
why $R_n$ is chosen as the default test and when it has the
best performance.

\section{Optimal Weights for Minimum Variance}
\label{s:optim-weights}

\subsection{Intuitions and definition}
\label{ss:geo}

The discussion thus far has been mainly focused on
understanding $U_{(i)}$ as a pivotal quantity,
that is, its distribution under the null hypothesis.
To help see what evidence in $Y_i$'s
against the null hypothesis,
denote by $x_{(i)}$ the $\mu_i$-th quantile
of $F^*(x)$, that is, $F^*(x_{(i)})=\mu_i$.
\ifthenelse{1=1}{
For a simple approximation, we write $Y_i$ in \eqref{eq:AD-02} as follows.
\[Y_i = Y_i^{(0)} + D_i\]
where
\[ 
Y_i^{(0)} = -2\replace{}{\mu_i(1-\mu_i)}\left[\mu_i\ln\frac{F^*(X_{(i)})}{\mu_i} +(1-\mu_i)\ln\frac{1-F^*(X_{(i)})}{1-\mu_i}\right]
\] 
and
\begin{equation}\label{eq:D-i}
D_i =  -2\replace{}{\mu_i(1-\mu_i)}\left[\mu_i\ln\frac{F(X_{(i)})}{F^*(X_{(i)})} +(1-\mu_i)\ln\frac{1-F(X_{(i)})}{1-F^*(X_{(i)})} \right].
\end{equation}
The expected evidence is seen to be the expected value of $D_i$
because $Y_i^{(0)}$  has the same distribution as $Y_i$ when $Y_i$ is under the null hypothesis.
For a simple approximation to the quantity $D_i$ in \eqref{eq:D-i}, we use its Taylor expansion at $F(X_{(i)}) = F^*(X_{(i)})$ up to second order:
\begin{eqnarray*}
&&-2\replace{}{\mu_i(1-\mu_i)}\left[ \frac{\mu_i}{F^*(X_{(i)})} - \frac{1-\mu_i}{1-F^*(X_{(i)})}\right] \left[ F(X_{(i)}) - F^*(X_{(i)})\right]\\ &&
+\replace{}{\mu_i(1-\mu_i)}\left[\frac{\mu_i}{[F^*(X_{(i)})]^2} + \frac{1-\mu_i}{[1-F^*(X_{(i)})]^2}\right] \left[ F(X_{(i)}) - F^*(X_{(i)})\right]^2
\end{eqnarray*}
To see the expected evidence, we take the following approximation by replacing the sampling quantile $X_{( i)}$ with the corresponding theoretical quantile $x_{(i)}$
\begin{equation}\label{eq:S2N}
\replace{}{\mu_i(1-\mu_i)}\left[\frac{1}{\mu_i} + \frac{1}{1-\mu_i}\right] \left[ F(x_{(i)}) - F^*(x_{(i)})\right]^2 
        \;=\; \replace{\frac{\left[ F(x_{(i)}) - F^*(x_{(i)})\right]^2}{\mu_i(1-\mu_i)}.}
        {
        \left[ F(x_{(i)}) - F^*(x_{(i)})\right]^2.
        }
\end{equation}
Combining the approximations \eqref{eq:AD-02} and \eqref{eq:S2N}, we have the following approximation.
\begin{equation}\label{eq:evidence-01}
	\replace{E\left(Y_i\right) - \frac{1}{n+2}
	\approx 
	\frac{\left[F(x_{(i)}) - F^*(x_{(i)})\right]^2}{\mu_i(1-\mu_i)},}
 {
E\left(Y_i\right) - \frac{\mu_i(1-\mu_i)}{n+2}
	\approx 
	\left[F(x_{(i)}) - F^*(x_{(i)})\right]^2, 
 }
\end{equation}
for $i=1,...,n$, when the sample $X_1,..., X_n$ has been drawn from $F^*(x)$. Clearly, the expected evidence 
\eqref{eq:evidence-01} has an attractive interpretation as 
a type of signal-to-noise ratio.

}{
Then
\begin{equation}\label{eq:evidence-01}
	E\left(Y_i\right) - 1
	{\color{red}\approx} 
	\frac{\left[F(x_{(i)}) - F^*(x_{(i)})\right]^2}{\mu_i(1-\mu_i)},
\end{equation}
a type of signal-to-noise ratio.
}
This implies that the weighted Anderson-Darling $R_n^2(w)$
deals with the $n$ pieces of unknown quantities 
$\left[F(x_{(i)}) - F^*(x_{(i)})\right]^2\replace{/[\mu_i(1-\mu_i)]}{}$
by capturing their weighted sum:
\begin{equation}\label{eq:evidence-02}
E\left[R_n^2(w)\right] - \frac{1}{n+2}\sum_{i=1}^nw_i\replace{}{\mu_i(1-\mu_i)}
	\approx 
\sum_{i=1}^nw_i \replace{\frac{\left[F(x_{(i)}) - F^*(x_{(i)})\right]^2}{\mu_i(1-\mu_i)}.}
{
\left[F(x_{(i)}) - F^*(x_{(i)})\right]^2
}
=\sum_{i=1}^nw_i\delta_i,
\end{equation}
where
\begin{equation}\label{eq:delta-01}
	\replace{ \delta_i =  \frac{\left[F(x_{(i)}) - F^*(x_{(i)})\right]^2}{\mu_i(1-\mu_i)}
  {\color{red} \qquad ? \delta_i =  \left[F(x_{(i)}) - F^*(x_{(i)})\right]^2}}
  {
  \delta_i =  \left[F(x_{(i)}) - F^*(x_{(i)})\right]^2
  }
\end{equation}
for $i=1,...,n$. Therefore, for small values of $\delta_i$'s in \eqref{eq:delta-01},
the corresponding noise level in capturing the signal $\sum_{i=1}^n w_i\delta_i$ in \eqref{eq:evidence-02}
is determined by the variance of $R_n^2(w)=\sum_{i=1}^n w_iY_i$, that is,
\begin{equation}\label{eq:var-Y-01}
w'\Sigma w= 
w'\mbox{Cov}\left(Y, Y'\right) w	
\end{equation}
under the null hypothesis, where $w=(w_1,...,w_n)'$, $Y=(Y_1,...,Y_n)'$, and $\Sigma=\mbox{Cov}\left(Y, Y'\right) $ denotes the variance and covariance matrix of $Y$.
This implies that for a given direction of the distributional deviations $\delta_i$'s, we can find the optimal weights and that for a given vector of weights $w$, we have a unique direction of the distributional deviations for which the test statistic $R_n^2(w)$ allows good performance to be achieved. We summarize this observation more precisely in the following lemma.

\begin{lemma}\label{lemma:weights-and-direction}
Let $\delta=(\delta_1,...,\delta_n)'$, where $\delta_i$ is defined in
\eqref{eq:delta-01}.
Denote by $\Sigma$ the covariance matrix of $Y=(Y_1,...,Y_n)'$,
	where $Y_i$ is defined in \eqref{eq:AD-02}.
	If $\delta\neq 0$, then
	for all $w\neq 0$
\begin{eqnarray*}
	\frac{(w'\delta)^2}{w'\Sigma w}
	\leq \delta\Sigma^{-1}\delta
\end{eqnarray*}
with the equality hold if and only if $w \propto \Sigma^{-1}\delta.$
\end{lemma}
\color{black}

This is a familiar mathematical result and
can be proved straightforwardly by applying the Cauchy-Schwarz inequality
theorem to the two vectors $\Sigma^{\frac{1}{2}}w$ and 
$\Sigma^{-\frac{1}{2}}\delta$.
The covariance matrix of $Y=(Y_1,...,Y_n)'$
is given in Theorem \ref{thm:exact-cov} below in Section
\ref{ss:finite-cov}.
This allows for a geometric interpretation regarding the performance of
the test statistic $R_n^2(w)$:
for any given vector of weights $w$,
the test statistic is the most powerful
when the direction $\delta$
is proportional to $\Sigma w$.
We call the direction $\Sigma w$ the focal direction of the weighted test
$R_n^2(w)$.

\begin{figure}[!htb]
\centering
\includegraphics[width=6.0in]{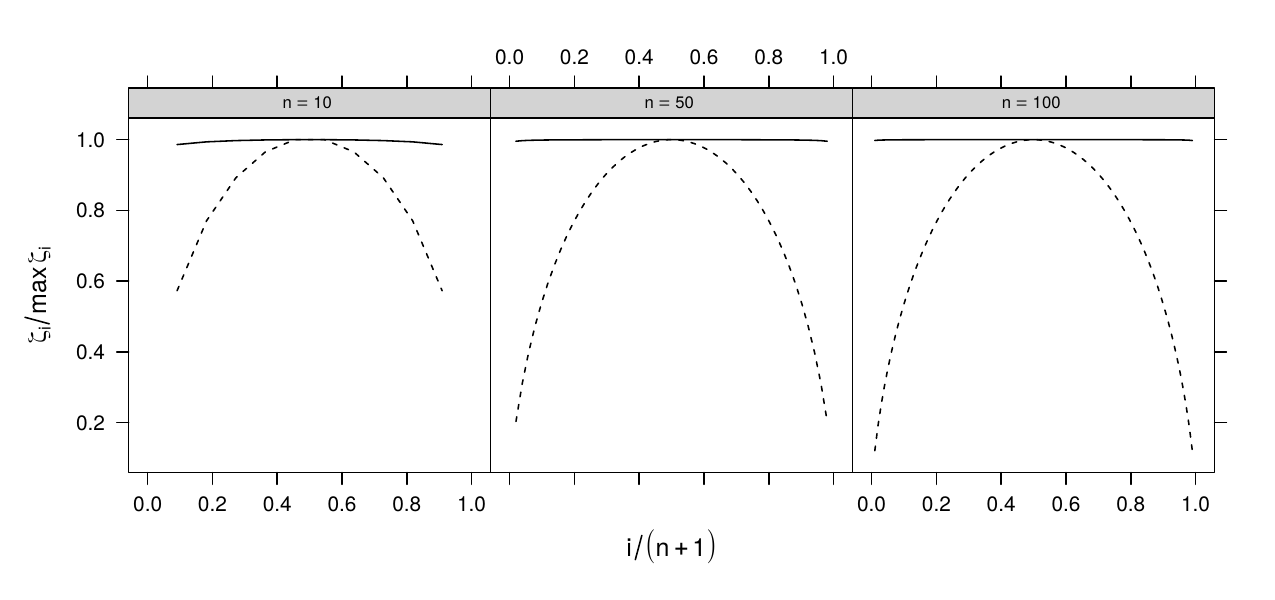}
        \caption{The variance-adjusted focal directions of
        Anderson-Darling test statistic (dashed line) and the test statistic with
 $w_i = 1/[\mu_i(1-\mu_i)]^2$ (solid line), defined in Subsection \ref{ss:geo} with elements \eqref{eq:var-adjusted-direction-01},
        for $n=10$, $50$, and $100$.
        }
\label{fig:AD-focal-direction-01}
\end{figure}

Taking into account the variance of $U_{(i)}$ as a measure of uncertainty, here we also use the variance adjusted direction
$\zeta$ with its $i$-th component defined as
\begin{equation}\label{eq:var-adjusted-direction-01}
\zeta_i = \frac{\delta_i}{\mu_i(1-\mu_i)},\qquad i=1,...,n.
\end{equation}
Using the variance-adjusted direction in \eqref{eq:var-adjusted-direction-01}, the variance-adjusted focal direction of Anderson-Darling
is \[\replace{$\delta\propto \Sigma \mathbf{1}$, }{
\zeta\propto \mbox{diag}\left(\frac{1}{\mu_1(1-\mu_1)},...,\frac{1}{\mu_n(1-\mu_n)}\right)\Sigma \left(\frac{1}{\mu_1(1-\mu_1)},...,\frac{1}{\mu_n(1-\mu_n)}\right)',}\]
which 
 is shown in Figure \ref{fig:AD-focal-direction-01}
for the three cases of $n=10$, $50$, and $100$,
where $\mbox{diag}(x)$ stands for the diagonal matrix with its argument vector $x$ as the diagonal elements.
Clearly,  
due to the strong correlations among $U_{(i)}$'s,
Anderson-Darling effectively focuses on the central area more than
the tail areas. 

So far, it has been seen that
the challenge of goodness-of-fit is due to
the fact that it essentially involves simultaneous inference on multiple
parameters $F^*(X_{(i)})$, $i=1,...,n$.
Therefore, in general, there are no uniformly optimal weights.
Consequently, we can only consider optimal weights
by considering a certain type of performance,
such as typical or average performance.
Consider the case where $\delta_i \propto \mu_i(1-\mu)$ are exchangeable. 
The use of $\delta_i \propto \mu_i(1-\mu)$ reflects the consideration that the variance of $U_{(i)}$
is $\mu_i(1-\mu_i)/(n+2)$.
This can be interpreted in practice as a type of average
from experiment to experiment.
By {\it exchangeable}, we mean that there is a Bayesian type of explanation, {\it that is}, the distribution of $\delta_i$'s, obtained from experiment to experiment, is invariant with respect to the permutations of $\delta_i$'s.
It follows from Lemma \ref{lemma:weights-and-direction} that
the optimal weights $w$ in this case are the optimal weights
that minimize the variance of $R_n^2(w)$.
Formally,
for the sake of terminology convenience,
we define optimal weights for minimum variance as follows.
\begin{definition}\label{def:optim}
The weights $w^{\mbox{(optim)}}$ satisfying 
	\replace{$\sum_{i=1}^nw_i=c$}{$\sum_{i=1}^nw_i\mu_i(1-\mu_i)=c$} for some positive $c$ are called 
optimal for minimum variance if
\[ w^{\mbox{(optim)}} = \arg\min_{\sum_{i=1}^nw_i\mu_i(1-\mu_i)=c}
\mbox{Var}\left( R_n^2(w)\right).
\]
\end{definition}

We discuss the optimal weights below in Sections \ref{ss:finite-cov}
and \ref{ss:limit-cov} for the finite-sample and large-sample cases, respectively.

\subsection{Finite sample cases}
\label{ss:R2n}
\label{ss:finite-cov}

The optimal weights $w_{\mbox{opt}}$ depend on the evaluation of
the variance of $R_n^2(w)$. 
Since $R_n^2(w)$ is linear in $\ln\left(U_{(i)}\right)$'s and 
$\ln\left(1-U_{(i)}\right)$'s, 
 the variance of $R_n^2(w)$ is given by
\[
4\sum_{i=1}^n w_i\sum_{j=1}^nw_j\mbox{Cov}\left(
B_i,
B_j
\right)
\]
with 
$
B_i = \mu_i^2(1-\mu_i)\ln (U_{(i)})+\mu_i(1-\mu_i)^2\ln(1-U_{(i)})
$ and
\begin{eqnarray*}
&& \mbox{Cov}\left(
B_i,
B_j
\right)\\
&=&\scriptsize
(\mu_i^2(1-\mu_i), \mu_i(1-\mu_i)^2)\left[\begin{array}{cc} 
\mbox{Cov}(\ln (U_{(i)}), \ln (U_{(j)})) &\mbox{Cov}(\ln (U_{(i)}),\ln(1-U_{(j)}))
\\
\mbox{Cov}(\ln (1-U_{(i)}), \ln (U_{(j)})) &\mbox{Cov}(\ln (1-U_{(i)}),\ln(1-U_{(j)}))\end{array}
\right]\left(\begin{array}{c} \mu_j^2(1-\mu_j)\\ \mu_j(1- \mu_j)^2\end{array}\right).
\end{eqnarray*}
The entries in the above $2\times 2$ matrix are given in 
the following theorem.

\begin{thm}
\label{thm:exact-cov}
Let $U_{(1)} < ... < U_{(n)}$ be the sorted uniforms of size $n$.
Denote by $\psi(x)$ and $\psi_1(x)$
the digamma and trigamma functions, respectively.
Then 
\begin{enumerate}
\item[\rm (a)] $\mbox{E}\left[\ln\left(U_{(i)}\right)\right]
		=\psi(i)-\psi(n+1)$,
$\mbox{E}\left[\ln\left(1-U_{(i)}\right)\right] =\psi(n+1-i)-\psi(n+1)$,
$\mbox{Var}\left[\ln\left(U_{(i)}\right)\right] =\psi_i(i)-\psi_i(n+1)$, and
$\mbox{Var}\left[\ln\left(1-U_{(i)}\right)\right] =\psi_1(n+1-i)-\psi_1(n+1)$
		for $i=1,...,n$;
\item[\rm (b)]
$\mbox{Cov}\left[\ln\left(U_{(i)}\right),\; \ln\left(U_{(j)}\right)\right] =\psi_1(j)-\psi_1(n+1)$
		and
$\mbox{Cov}\left[\ln\left(1-U_{(i)}\right),\; \ln\left(1-U_{(j)}\right)\right] =\psi_1(n+1-i)-\psi_1(n+1)$
		for all $1\leq i<j\leq n$;
		and
\item[\rm (c)]
$\mbox{Cov}\left[\ln\left(U_{(i)}\right),\; \ln\left(1-U_{(j)}\right)\right] =-\psi_1(n+1)$
	and
\begin{eqnarray}
	\mbox{Cov}\left[ \ln(1-U_{(i)}),\;\ln(U_{(j)}) \right]
	&=&  \frac{\Gamma(n+1)}{\Gamma(i)}
	\sum_{k=1}^\infty
	\frac{\Gamma(i+k)}{k\Gamma(n+1+k)}
	\left[\psi(n+1+k)-\psi(j+k)\right]
	\nonumber\\
	&&
	-
	\left[\psi(n+1-i)-\psi(n+1)\right] \left[\psi(j)-\psi(n+1)\right]
	\nonumber 
\end{eqnarray}
		for all $1\leq i<j\leq n$.
\end{enumerate}
\end{thm}

\begin{figure}[!htb]
\centering
\includegraphics[width=6.0in]{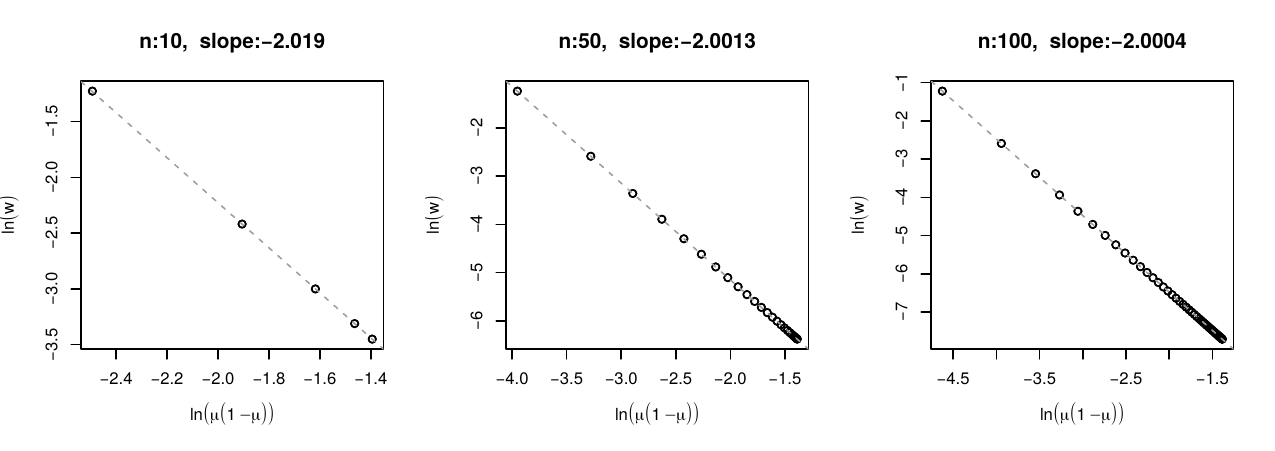}
        \caption{The exact finite-sample results on optimal weights
        for three cases $n=10$, 50, and 100.
        The slope is the regression coefficient of the least-squares fit of
        the optimal weights on the variance of $U_{(i)}$, both on logarithmic scale.
        }
\label{fig:weights-finite-sample-01}
\end{figure}

The proof of Theorem \ref{thm:exact-cov} is provided in Appendix \ref{ss:proof-exact-cov}. 
The optimal weights for $n=10$, $50$, and $100$
are shown in Figure \ref{fig:weights-finite-sample-01}.
These numerical results clearly suggest that optimal weights are 
almost 
inversely proportional to the variances of $U_{(i)}$, that is,
\begin{equation}\label{eq:weights-opt}
	w_i \propto \frac{1}{\mu_i^2(1-\mu_i)^2} = 
	\frac{(n+1)^4}{i^2(n+1-i)^2}
\end{equation}
for $i=1,...,n$.
These optimal results are, in fact, asymptotically exact. This 
is discussed in the next subsection.
The numerical results in Figure \ref{fig:weights-finite-sample-01}
also show that for small $n$, the optimal weights weight slightly more on the tails
of $F(x)$, but they appear to converge very quickly to \eqref{eq:weights-opt}.

\subsection{Large sample results}
\label{ss:limit-cov}

Since the variance $\mbox{Var}\left(U_{(i)}\right)=\frac{1}{n+2}\mu_i(1-\mu_i)$
converges to zero as $n\rightarrow \infty$,
we use the delta method to approximate $R_n^2(w)$
in terms of $U_{(i)}$'s.  This allows us to work conveniently with 
$U_{(i)}$'s for investigating the 
large-sample behavior of $R_n^2$.  Since the coefficient
of the corresponding Taylor expansion for the first order
$U_{(i)}-\mu_i$ is zero,
we have
\begin{equation}
\label{eq:RnW-approx-03}
	R_n^2(w) \approx 
  \sum_{i=1}^n \replace{\frac{w_i}{\mu_i(1-\mu_i)}}{w_i} (U_{(i)} - \mu_i)^2.
\end{equation}
It is seen that in this case, to find the optimal weights
for the approximation in \eqref{eq:RnW-approx-03},
{\it i.e.},
\[
	r_n^2(w) = 
	\sum_{i=1}^n \replace{\frac{w_i}{\mu_i(1-\mu_i)}}{w_i} (U_{(i)} - \mu_i)^2,
	\]
we need to work with the covariance matrix of the vector of squared
$\left(U_{(i)}-\mu_i\right)$'s.
The results on this covariance matrix are summarized into the following theorem, with
the proof given in Appendix \ref{app:cov}.

\begin{thm}
\label{thm:finite-cov}
Let $U_{(1)} < ... < U_{(n)}$ be the sorted uniforms of size $n$.
Then
\begin{eqnarray*}
		&&
	\mbox{Cov}\left[ (U_{(i)}-\mu_i)^2,\;
		(U_{(j)}-\mu_j)^2 \right] \\
		&=&
 \frac{2\mu_i^2 (1-\mu_j)^2}{(n+2)(n+3)}
 + \frac{\mu_i (1-\mu_j)}{(n+2)(n+3)}
	\left\{ \frac{3(1-3\mu_i)(2-3\mu_j)}{n+4}
	- \frac{(1-\mu_i)\mu_j}{(n+2)}
	\right\}
\end{eqnarray*}
hold for all $1\leq i\leq j\leq n$.
\end{thm}

It is not difficult to see that as $n\rightarrow \infty$,
the scaled covariance
$n^2 \mbox{Cov}\left[ (U_{(i)}-\mu_i)^2,\; (U_{(j)}-\mu_j)^2 \right]$
converges to $2\mu_i^2 (1-\mu_j)^2$.
This is closely related to the mathematical 
treatment of \cite{anderson1952asymptotic}
using the limiting process of the uniform empirical process 
$\sqrt{n}\left[G_n(u)-u\right]$, $u\in [0,1]$, where $G_n(u)$ denotes the empirical distribution derived from $u_1 = U_{(1)},..., u_n=U_{(n)}.$
Here, we use the uniform quantile process defined on $[0,1]$
\citep{csorgo1978strong, shorack1972functions},
with a minor modification that replaces $\sqrt{n}$ with $\sqrt{n+2}$,
\begin{equation}\label{eq:unif-quan-proc}
 \mathcal{B}_n(t) = \sqrt{n+2}[G^{-1}_n(t)-t],\qquad (0\leq t\leq 1)
\end{equation}
Note that the inverse $G^{-1}_n(t)$ of $G_n(u)$ will always be the left continuous one.
Intuitively, the uniform quantile process is simply a continuous extension of the 
sequence
\[\mathcal{B}_n(i/(n+1)) = \sqrt{n+2}[U_{(i)}-E(U_{(i)})], \qquad
(i=1, ..., n).\]
It is known that the limiting process of the uniform quantile process is also the Brownian bridge \citep{csorgo1978strong, shorack1972functions}.
Therefore, applying Slutsky's theorem, we see that the limiting process of $\mathcal{B}_n(t)$
is also the Brownian bridge.
In this case, we can easily obtain the corresponding results
for the covariance structure of the limiting process, which is characterized
in the following theorem. 
\begin{thm} 
\label{thm:limit-cov}
The limiting process of $\mathcal{B}_n(t)$ defined in \eqref{eq:unif-quan-proc} 
is the Brownian bridge. For the Brownian bridge,
\[
\mbox{Cov}\left[ \mathcal{B}(s),\; \mathcal{B}(t) \right] = s(1-t)
\]
and
\[
\mbox{Cov}\left[ \mathcal{B}^2(s),\; \mathcal{B}^2(t) \right] = 2s^2(1-t)^2
\]
holds  for all $ 0\le s\leq t\leq 1$. 
\end{thm}

The proof is given in Appendix \ref{app:limit-cov}.
From this result, we can obtain the asymptotic optimal
weights or, more exactly, the optimal weight function.
The result is summarized into the following theorem, with
the proof given in Appendix \ref{app:symp-optim}.

\begin{thm}\label{thm:aymptotic-weights}
In the limit as $n\rightarrow\infty$,
the optimal weight function is given by
\begin{equation}\label{eq:asym-optim-weights}
	\psi(t) \propto 
	\frac{1}{[t(1-t)]^2} \qquad(t \in [\varepsilon, 1-\varepsilon]),
\end{equation}
for all $\varepsilon \in\left(0, \frac{1}{2}\right]$.
\end{thm}

This result provides theoretical support for
the use of the weights defined in
\eqref{eq:opti-weight-00} as the optimal weights in practice
for all sample sizes $n$.

\section{Circularly Symmetric Tests}
\label{s:cs}

\subsection{The circular process of uniform spacing on the unit circle}
\label{ss:circular-process}

Since the distributional deviations in the central areas of $F(x)$ captured 
by the $U_{(i)}$'s are averaged together with those in the tail areas, the
efficiency of methods with test statistics defined through the $U_{(i)}$'s is questionable.
All of these methods suffer from the confounded effect of different
locations and various signal frequencies in the deviations
from the null hypothesis.
This motivates the idea of the following circularization to eliminate the location effect, 
so that we allow the weights to focus on the various signal frequencies.

\begin{figure}[!htb]
\centering
        \includegraphics[width=4.0in]{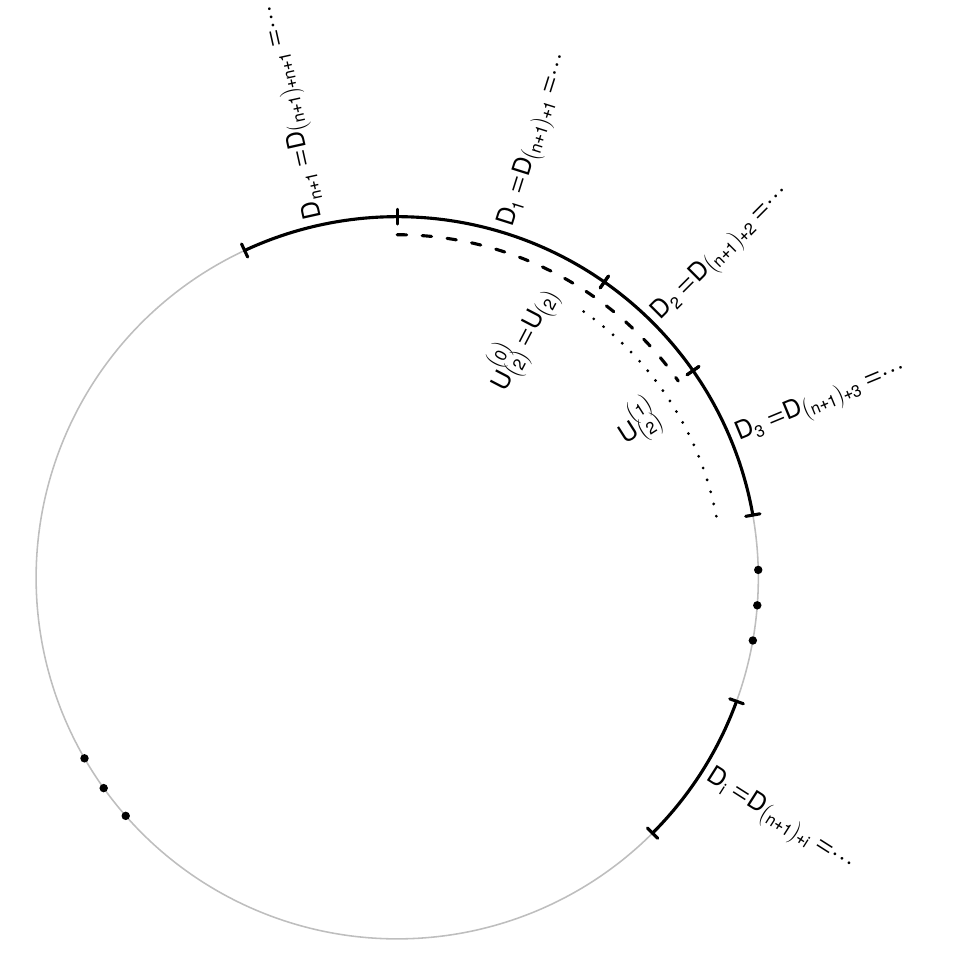}
        \caption{
                The circular uniform spacings defined in \eqref{eq:XDi},
                $D_{k(n+1)+i} = D_i = U_{(i)}-U_{(i-1)}$ for
                $i=1, ..., n+1$ and all $k=0, 1, ...$. Two examples of the circular countparts of $U_{(i)}$,
  {\it i.e.}, $U_{(i)}^{(c)}$ defined in \eqref{eq:CS-02}, are shown by the dashed and dotted circular arcs for $U_{(2)}^{(0)}$ and $U_{(2)}^{(1)}$.
        }
\label{fig:circular-uniform-spacing}
\end{figure}

The circularization technique is straightforward. To set the stage,
let $U_{(0)}=0$ and let $U_{(n+1)}=1$.
Extend the uniform spacings 
\[ 
D_i=U_{(i)}-U_{(i-1)}, \qquad i=1, ..., n+1,
\] 
as
a circular process at $n+1$ locations,
as depicted in Figure \ref{fig:circular-uniform-spacing}.
In this case, $D_i$'s can also be seen as the uniform spacings on the unit circle.
Formally, the extended $D_i$'s 
are defined as 
$D_i = D_{i \mbox{ mod } (n+1)}$, $i=1,2,...,$ that is,
\begin{equation}\label{eq:XDi}
	D_{k(n+1)+i}
 = D_i \qquad \mbox{for $i=1, ..., n+1$ and all $k=0, 1, ...$.}
\end{equation}
Accordingly, we also extend the definition of $U_{(i)}$'s as

\begin{equation}\label{eq:XUi}
U_{(i)} = \sum_{j=1}^{i} D_j\qquad \mbox{for $i=1, 2, ...$}
\end{equation}
where $D_i$'s are given in \eqref{eq:XDi}.

Note that the statistics $U_{(i)}$'s are the
cumulative sums of the uniform spacings $D_i$'s, starting from $D_1$. 
We define the circular counterparts of $U_{(i)}$
as the cumulative sums of the uniform spacings $D_i$'s but starting from $D_{c+1}$
for $c=0,1,...,n$:
\begin{equation}\label{eq:CS-02}
	U^{(c)}_{(i)} = U_{(c+i)} - U_{(c)}
	=\sum_{k=c+1}^{c+i} D_k
	\qquad(c=0,...,n; i=1,...,n+1)
\end{equation}
where $U_{(i)}$'s are given in \eqref{eq:XUi}. Note that $U_{(i)} = U_{(i)}-U_{(0)}$ is valid for all $i=1,...,n.$
The class of statistics \eqref{eq:CS-02} contains the original $U_{(i)}$ 
as the special case with $c=0$.

\subsection{The method of circularization}
Let $T_n=T_n(U_{(1)}, ..., U_{(n)})$ be any test statistic created using
$U_{(1)} = F(X_{(1)}), ..., U_{(n)} = F(X_{(n)})$, where $F(.)$ is the distribution function under the null hypothesis. We assume that large values of $T_n$ provide evidence against the null hypothesis.
The class of such test statistics includes all the test statistics considered in this paper. Here we propose
a method of circularization to construct circularly symmetric version of $T_n.$ 

Our proposed circularization method consists of two steps: a {\it looping} step and a {\it pooling} step. 
The looping step runs for $c=0, 1, ...,n$ to define the $n+1$ statistics
\begin{equation}\label{eq:looping-step}
    T_{n, c}
    = T_n\left(U^{(c)}_{(1)}, ..., U^{(c)}_{(n)}\right),\qquad c=0, 1, ...,n,
\end{equation}
where $U^{(c)}_{(1)}, ..., U^{(c)}_{(n)}$ are given in \eqref{eq:CS-02}. The pooling step 
pools the evidence in \eqref{eq:looping-step}. 
The resulting statistic is called the pooled or circularized test statistic and is denoted by
\begin{equation}\label{eq:pooling-step}
   T_n^{(\mbox{pooled})} = \mbox{Pool}\left(T_{n,0}, T_{n,1}, ..., T_{n,n}\right).
\end{equation}
In this paper, we consider two types of pooling operations: the {\it average} pooling and
the {\it max} pooling. The average pooling and the max pooling are defined as
\begin{equation}
\label{eq:CS-mean-T}
    T_n^{(\mbox{avg})}
    = \frac{1}{n+1}\sum_{c=0}^{n} T_{n,c}
\end{equation}
and
\begin{equation}
\label{eq:CS-max-T}
    T_n^{(\mbox{max})}
    = \max_{c\in\{0,...,n\}} T^{(c)}_n,
\end{equation}
respectively. Thus, like their parent test statistics, large values of the pooled test statistics provide evidence against the null hypothesis.

\begin{remark}
    It is easy to see that both $T_n^{(\mbox{avg})}$ and $T_n^{(\mbox{max})}$, defined in
    \eqref{eq:CS-mean-T} and \eqref{eq:CS-max-T}, are circularly symmetric with respect to the circular process introduced in Section \ref{ss:circular-process} because all possible $T_{n,c}$'s on the circle are considered and the two pooling operations are invariant with respect to permutations of $T_{n,c}$'s.
\end{remark}

\begin{remark}
  Note that the looping step takes a fixed-size $(n+1)$ sliding window of indices of $D_i$ with the starting index $c=0, 1, ...,n$ and obtains the test statistic of interest using all the $D_i$'s in each window. Thus, in an analogy with the concept of scan statistic \citep{naus1965distribution}, we may call
  the statistic $T_{n,c}$ defined in \eqref{eq:looping-step} a scan statistic.
\end{remark}

The detailed construction of the two types of circularization for the test statistics $W_n^2$ and $R_n^2$ is provided as follows.
For the test statistic $W_n^2$, the looping step 
defines
\[
W_{n,c}^2 = -2 \sum_{i=1}^n \left[\mu_i
	\ln \frac{U^{(c)}_{(i)}}{\mu_i} + (1-\mu_i)\ln
	\frac{\left(1-U^{(c)}_{(i)}\right)}{1-\mu_i}
	\right],\qquad c=0,1,...,n.
\]
This leads to the two circularized test statistics  
\begin{equation}\label{eq:normalized-Wn}
	\tilde{W}_n^2 \equiv \frac{1}{n+1}\sum_{c=0}^{n} W_{n,c}^2
  = -\frac{2}{n+1}	
	\sum_{c=0}^n\sum_{i=1}^n
	\left[\mu_i\ln\frac{U_{(i)}^{(c)}}{\mu_i}
	+(1-\mu_i)\ln\frac{1-U_{(i)}^{(c)}}{1-\mu_i}
	\right]
\end{equation}
and
\begin{equation}\label{eq:normalized-Wn-vee}
	\overset{_{_{_\vee}}}{W_n^2} \equiv \max_{c\in\{0,...,n\}} W_{n,c}^2
 = \max_{c\in\{0,...,n\}}\left(-2
	\sum_{i=1}^n
	\left[\mu_i\ln\frac{U_{(i)}^{(c)}}{\mu_i}
	+(1-\mu_i)\ln\frac{1-U_{(i)}^{(c)}}{1-\mu_i}
	\right]\right),
\end{equation}
using the average pooling and the max pooling, respectively.
Similarly, for the test statistic $R_n^2$, the pooling step 
gives
\[
R_{n,c}^2 = - \frac{1}{\sum_{k=1}^n\frac{1}{k}}\sum_{i=1}^n \left[\frac{1}{1-\mu_i}
	\ln \frac{U^{(c)}_{(i)}}{\mu_i} + \frac{1}{\mu_i}\ln
	\frac{\left(1-U^{(c)}_{(i)}\right)}{1-\mu_i}
	\right],\qquad c=0,1,...,n.
\]
The pooling step gives the two circularized test statistics
\begin{equation}\label{eq:normalized-Rn}
	\tilde{R}_n^2 \equiv 
 \frac{1}{n+1}\sum_{c=0}^{n} R_{n,c}^2 
 = - 
	\frac{1}{(n+1)\sum_{k=1}^n\frac{1}{k}}	
	\sum_{c=0}^n\sum_{i=1}^n
	\replace{\frac{1}{\mu_i(1-\mu_i)}
	\left[\mu_i\ln\frac{U_{(i)}^{(c)}}{\mu_i}
	+(1-\mu_i)\ln\frac{1-U_{(i)}^{(c)}}{1-\mu_i}
	\right].}{
 \left[\frac{1}{1-\mu_i}\ln\frac{U_{(i)}^{(c)}}{\mu_i}
	+\frac{1}{\mu_i}\ln\frac{1-U_{(i)}^{(c)}}{1-\mu_i}
	\right]
    }
\end{equation}
and
\begin{equation}\label{eq:normalized-Rn-vee}
	\overset{_{_{_\vee}}}{R_n^2} \equiv \max_{c\in\{0,...,n\}} R_{n,c}^2
 =\max_{c\in\{0,...,n\}}\left(
         - 
	\frac{1}{\sum_{k=1}^n\frac{1}{k}}	
	\sum_{i=1}^n
	\replace{\frac{1}{\mu_i(1-\mu_i)}
	\left[\mu_i\ln\frac{U_{(i)}^{(c)}}{\mu_i}
	+(1-\mu_i)\ln\frac{1-U_{(i)}^{(c)}}{1-\mu_i}
	\right].}{
	\left[\frac{1}{1-\mu_i}\ln\frac{U_{(i)}^{(c)}}{\mu_i}
	+\frac{1}{\mu_i}\ln\frac{1-U_{(i)}^{(c)}}{1-\mu_i}
	\right]
    }\right)
\end{equation}
with the average pooling and max pooling, respectively.

It is easy to see that the above circularization technique can be applied to 
the original Anderson-Darling, Zhang's likelihood ratio,
Cram\'er-von Mises, and Kolmogorov-Smirnov test statistics.
In the next section, a simple simulation study is carried out
to compare the performance of these methods
and their circularized versions.
The large sample results for understanding and numerical
approximation are given in Section \ref{s:asym}
for $\tilde{W}_n^2$ and $\tilde{R}_n^2$.
The large sample results for $\overset{_{_{_\vee}}}{W_n^2} $
and $\overset{_{_{_\vee}}}{R_n^2}$
appear challenging and are expected to be considered
elsewhere.

\section{Power Comparison: a Simulation Study}
\label{s:numerical}
\label{s:power-comparison}


For power comparison, we focus on our investigation on 
a class of situations where the distributional deviations of $F^*(.)$ from
$F(.)$ are locally smooth but at different locations.
For this, we use, without loss of generality, the standard uniform $\mbox{Unif}(0,1)$ as $F(.)$ and consider 
the class of $F^*(.)$'s
obtained with simple local perturbations.
More precisely, $F^*(.)$ is given by the
probability density function ({\it pdf})
\[ 
f_{\eta, \sigma, \tau}(x)
	= 1 + \tau{\mathbf 1}_{(\eta-\sigma, \eta]}(x)
	 - \tau{\mathbf 1}_{(\eta, \eta+\sigma)}(x)
	 \qquad(0<x<1)
\] 
where $\sigma>0$, $0\leq \eta-\sigma$,
	 $\eta+\sigma \leq 1$, $0\leq \tau \leq 1$,
and ${\mathbf 1}_A(x)$ is the indicator function of the subset $A$,
that is, ${\mathbf 1}_A(x)=1$ if $x\in A$ and ${\mathbf 1}_A(x)=0$
otherwise.
The corresponding $F^*(.)$ has {\it pdf}
\[ 
F_{\eta, \sigma, \tau}(x)
=\left\{
\begin{array}{lll}
	x,&& \mbox{if $p \in (0,\eta-\sigma]$};\\
	x + \tau(x-\eta+\sigma),
	&&
		\mbox{if $x \in (\eta-\sigma, \eta]$};\\
	x - \tau(x-\eta-\sigma),
	&&
		\mbox{if $x \in (\eta, \eta+\sigma)$};\\
	x,&& \mbox{if $x \in [\eta+\sigma, 1)$},
\end{array}
\right.
 \qquad(0<x<1)
\] 
with the inverse {\it cdf}
\[
F^{-1}_{\eta, \sigma, \tau}(p)
=\left\{
\begin{array}{ll}
	p,& \mbox{if $p \in (0,\eta-\sigma]$};\\
	p -\frac{\tau(p-\eta+\sigma)}{1+\tau},&
		\mbox{if $p \in (\eta-\sigma, \eta+\tau\sigma]$};\\
	p + \frac{\tau(p-\eta-\sigma)}{1-\tau},&
		\mbox{if $p \in (\eta+\tau\sigma, \eta+\sigma)$};\\
	p,& \mbox{if $p \in [\eta+\sigma, 1)$},
\end{array}
\right.
 \qquad(0<p<1).
\] 

In this simulation study, we consider six test statistics: $W_n^2$, $R_n^2$, Anderson-Darling (AD), likelihood ratio (LR) of \cite{zhang2002powerful}, Cram\'er-von Mises (CvM), and Kolmogorov (KS) tests. For each of these test statistics, we also consider its two circularized circularly-symmetric (CS) versions with the average and max pooling operations \eqref{eq:CS-mean-T} and \eqref{eq:CS-max-T}. 
In this simulation study, we take the significance level of $0.05.$ 
The distributions of the test statistics are computed via Monte Carlo approximations with Monte Carlo sample size $10,000$. These include computing the critical values using 
Mote Carlo samples from $\mbox{Uniform}(0, 1)$.

\begin{figure}[!htb]
\centering
\includegraphics[width=6.5in]{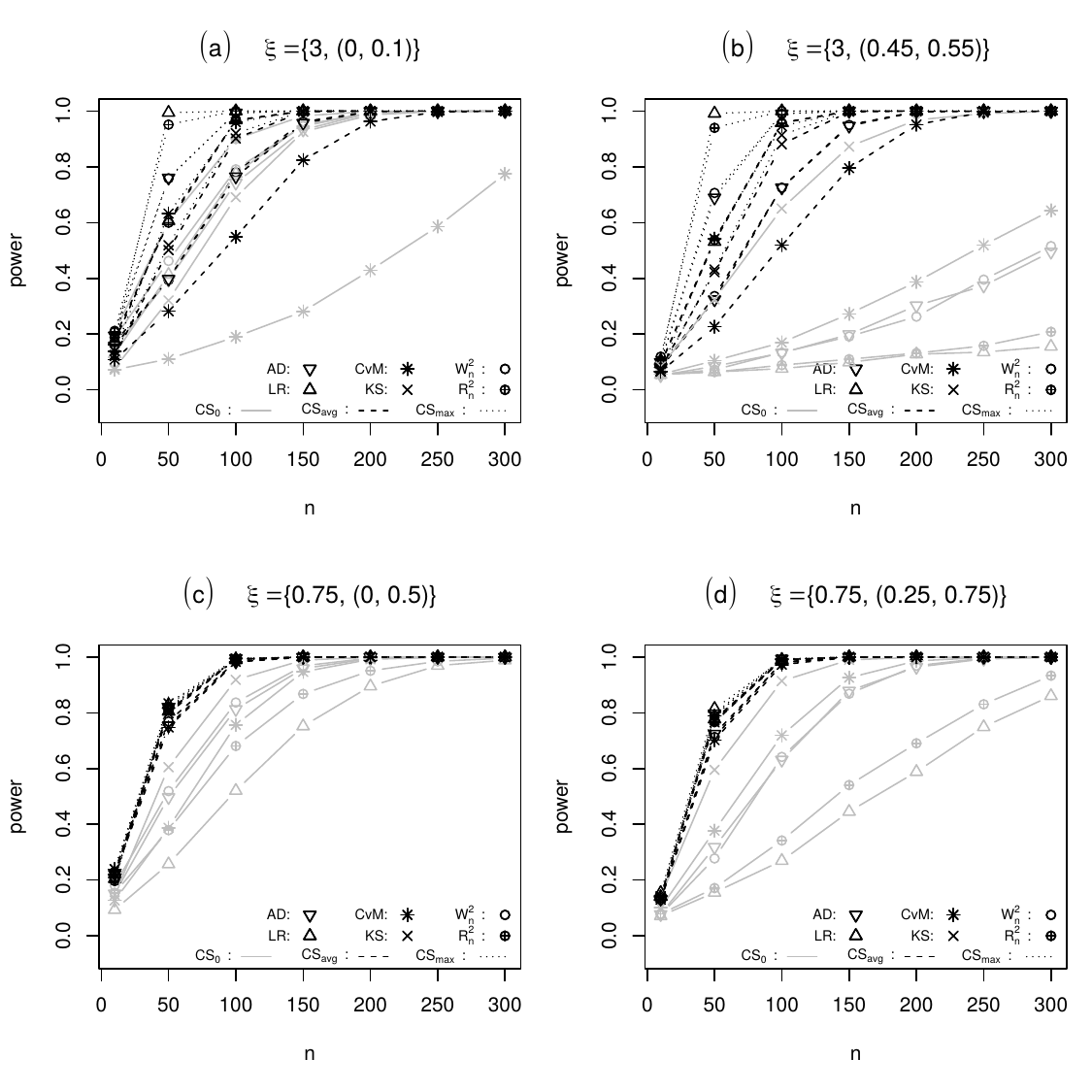}
        \caption{Power comparison in Section \ref{s:power-comparison} with significance level $0.05$.
        The title  $\xi=\{\tau, (\eta-\sigma, \eta+\sigma)\}$ of each plot
        represents the magnitude $\tau$ and
        location $(\eta-\sigma, \eta+\sigma)$
        where $F^*(.)$ deviates from $F(.)$, $\mbox{Uniform}(0, 1)$.
        The solid, dashed, and dotted curves correspond to $\mbox{CS}_0$ for the original tests,
        $\mbox{CS}_{\mbox{avg}}$ for the circularized version \eqref{eq:CS-mean-T}, and $\mbox{CS}_{\mbox{max}}$ for the circularized version \eqref{eq:CS-max-T}, respectively.
        }
\label{fig:pow-comp-01}
\end{figure}

For a simple but representative numerical comparison,
we consider two scenarios to be specified by the values of $\tau$, $\sigma$, and $\eta$ for $F_{\tau, \sigma, \eta}(x)$.
In the first scenario, we
have a small interval of length 0.1 with an appropriate
magnitude $\tau=3$.
The tail and central locations are specified by $\eta = 0.05$ and $0.5$.
The simulation results are tabulated in Tables \ref{tbl:pow-1} and  \ref{tbl:pow-2}
and shown in Plots (a) and (b) of Figure \ref{fig:pow-comp-01}, where for each test statistic,
$\mbox{CS}_0$ refers to the original or uncircularized version,
$\mbox{CS}_1=\mbox{CS}_{\mbox{avg}}$ the circularized version with the average pooling, and
$\mbox{CS}_2=\mbox{CS}_{\mbox{max}}$ the circularized version with the max pooling.
It is seen from Figure \ref{fig:pow-comp-01} (a) that all the methods have a good performance, expect for CvM, when the deviation of $F_{\tau, \sigma, \eta}(x)$ from $\mbox{Uniform}(0, 1)$ is in the tail areas. Figure \ref{fig:pow-comp-01} (b) shows that when the deviation of $F_{\tau, \sigma, \eta}(x)$ from $\mbox{Uniform}(0, 1)$ is in the central region, all the methods perform poorly, with KS performing slightly better. As expected, the circularized versions of all of the methods clearly eliminate this location effect. These results also show that the circularized versions of all the methods obtain dramatically improved performance.

In the second scenario, we
have a large interval of length 0.5 with an appropriate
magnitude $\tau=0.75$. Numerical results are tabulated in Tables \ref{tbl:pow-3} and  \ref{tbl:pow-4} and displayed in Plots (c) and (d) of Figure \ref{fig:pow-comp-01}. The results seem to suggest that
$\overset{_{_{_\vee}}}{W_n^2} $ and $\overset{_{_{_\vee}}}{R_n^2}$
improve the performance when signals are of high frequency,
while they are comparable to $\tilde{W}_n^2$ and $\tilde{R}_n^2$
for signals of low frequency.
One may reach the same conclusion as that in the first scenario. In summary, 
the results in Tables \ref{tbl:pow-1}--\ref{tbl:pow-4} and Figure \ref{fig:pow-comp-01} show clearly that the circularized LR method outperforms the circularized Anderson-Darling and that the circularized tests outperform their parent methods.

\section{Large-Sample Results}
\label{s:asym-dist}

\subsection{Weak Convergence}

In addition to the discussion on the weak convergence of $R_n^2$ in Section \ref{ss:limit-cov},
more discussion can be made based on the results in \cite{csorgo1988distributions} and
\cite{csorgHo1993convergence} and references therein.
For mathematical simplicity,
we define the uniform empirical quantile function on $[0,1]$ as
\[ 
        Q_n(0) = 0\quad\mbox{ and }\quad
Q_n(s) = U_{i,n}, \qquad \frac{i-1}{n+1} < s \leq \frac{i}{n+1}\qquad(i=1,...,n+1),
\] 
and the corresponding uniform empirical quantile process as
\[ 
        q_n(s) = \sqrt{n+1}\left[Q_n(s)-s\right], \qquad
        0\leq s\leq 1
\] 
and $Q_n(0) = 0$.
We consider the test statistic of the form
\begin{equation}\label{eq:ueq-test}
\frac{
        \int_{-\varepsilon} ^{1-\varepsilon} q_n^2(s) w(s) ds
        }{
        \int_{-\varepsilon} ^{1-\varepsilon} w(s) ds
        }
\end{equation}
where $0<\varepsilon<\frac{1}{2}$ and $w(s)>0$ on $[-\varepsilon, 1-\varepsilon]$.
It is easy to see that the difference between the uniform empirical quantile function $Q_n(s)$
defined above and that given in \cite{csorgo1988distributions},
\[
        U_n(0) = 0\quad\mbox{ and }\quad
U_n(s) = U_{i,n}, \qquad \frac{i-1}{n} < s \leq \frac{i}{n}\qquad(i=1,...,n),
\]
is small in the sense that
\[
        \sup_{s}\left|
        Q_n(s) - U_n(s)
        \right| \leq \frac{1}{n} \rightarrow 0,\qquad \mbox{as $n\rightarrow \infty$}.
\]
The weak convergence of \eqref{eq:ueq-test}, when
$Q_n(s)$ is replaced by $U_n(s)$, is given by
\cite{csorgo1988distributions}; see also
\cite{csorgHo1993convergence}.
By applying Slutsky's theorem, we can establish the weak convergence
of \eqref{eq:ueq-test}.
The weak convergence result also implies
the weak convergence of quadratic approximations to the test statistics $W_n^2$ and $R_n^2$ considered here
with the corresponding asymptotic distributions,
as they are Riemann sums of the path-wise integrals with
the mesh size of $1/(n+1)$, which goes to zero as $n$ goes to infinity.

The weak convergence of the circularized versions is subject to further rigorous study. The relevant results presented here are mostly heuristic, although the numerical results are consistent with the expected results.

\subsection{The Limiting Distribution of $R^2_n$}
\label{s:limiting-dist}

In this section, we investigate the asymptotic distribution of $R^2_n$ 
by taking the approach of \cite{anderson1952asymptotic}
and the extended result of their Theorem 4.1. In the present
case, the kernel function is
\[ \sqrt{w(s)}\sqrt{w(t)}[\min(s,t)-st]
=\frac{1}{s(1-s)} \frac{1}{t(1-t)} [\min(s,t)-st]
	\qquad (s, t \in [\varepsilon, 1-\varepsilon])
	\]
where $\varepsilon$ is a small positive number, say
$\varepsilon=1/[2(n+1)]$ in the context of a given sample.
We use a small $\varepsilon$ to rule out
index values near the two end points of the interval $(0,1)$
because the kernel function is unintegrable.
This does not mean we cannot consider the limiting distribution of $R^2_n$
for understanding the large-sample behavior of $R^2_n$
and for large-sample approximation to the distribution of $R^2_n$.
Theoretically, since $\varepsilon$ can be arbitrarily small, 
there is no problem to use the corresponding results for understanding of
the large-sample behavior of $R^2_n$.
Indeed, the results discovered below show that
the limiting distribution of  $R^2_n$
is a weighted sum of an infinite number of independently
squared standard normal random variables with
weights $1/\lambda_k$ decaying in a fashion proportional to
$1/k^2$.
Practically, for any finite sample of size of $n$,
the large-sample approximation to the distribution of $R^2_n$
is valid as long as it can provide satisfactory numerical approximations.
The use of $\varepsilon=1/[2(n+1)]$ is suggested
based on the fact that
when mapped into the interval $(0,1)$ as done in Section \ref{s:optim-weights},
the corresponding finite-sample extreme indices
are $1/(n+1)$ and $n/(n+1)=1-1/(n+1)$;
See also Remark \ref{remark:Cn}.
Alternative values can be used and are discussed below.

The next critical step is to solve the eigensystem defined by
the integral equation:
\begin{equation}\label{eq:eigensystem-01}
	f(t) = \lambda\int_{\varepsilon}^{1-\varepsilon}
	\sqrt{w(s)}\sqrt{w(t)}[\min(s,t)-st]f(s)ds.
\end{equation}
It is easy to find that
the solution satisfies the Sturm-Liouville equation
\citep[see, {\it e.g.},][]{anderson1952asymptotic}:
\begin{equation}\label{eq:Sturm-Liouville-0081}
	h''(t) + \lambda \psi(t) h(t)=0
\end{equation}
where $h(t) = f(t) \psi^{-\frac{1}{2}}(t)$.
This is known as an eigenvalue problem. The solution
can be found analytically and is summarized 
into the following theorem.

\begin{thm} \label{thm:limit-dist}
The solution to the integral equation \eqref{eq:eigensystem-01} is given by
a sequence of $\lambda_k = \omega_k^2 + \frac{1}{4}$ with the corresponding
eigenfunctions 
of two types. The first type is given by $\omega_k$s that satisfy
\begin{equation}\label{eq:omega-cos}
\tan\left(\omega_k\ln\frac{1-\varepsilon}{\varepsilon}\right)
	=\frac{1}{2\omega_k}
\end{equation}
and the corresponding eigenfunction
\[
	f_k(x) \propto \frac{1}{\sqrt{t(1-t)}} \cos\left(\omega_k\ln\frac{t}{1-t}\right)
	\qquad(t\in(\varepsilon, 1-\varepsilon)).
	\]
The second type is given by 
$\omega_k$s satisfying
\begin{equation}\label{eq:omega-sin}
\tan\left(\omega_k\ln\frac{1-\varepsilon}{\varepsilon}\right)
=-2\omega_k
\end{equation}
and the corresponding eigenfunction
\[
	f_k(x) \propto \frac{1}{\sqrt{t(1-t)}} \sin\left(\omega_k\ln\frac{t}{1-t}\right)
	\qquad(t\in(\varepsilon, 1-\varepsilon)).
	\]
Moreover, this solution corresponds to the Sturm-Liouville problem
with the Sturm-Liouville equation \eqref{eq:Sturm-Liouville-0081}
and the Robin boundary conditions 
	\[
		f(\varepsilon) -2(1-\varepsilon)f'(\varepsilon)=0
		\quad\mbox{ and }\quad
		f(1-\varepsilon) +2(1-\varepsilon)f'(1-\varepsilon)=0
		\]
and, thereby, all the eigenfunctions
are orthogonal to each other. 
\end{thm}

\begin{figure}[!htb]
\centering
\includegraphics[width=6.0in]{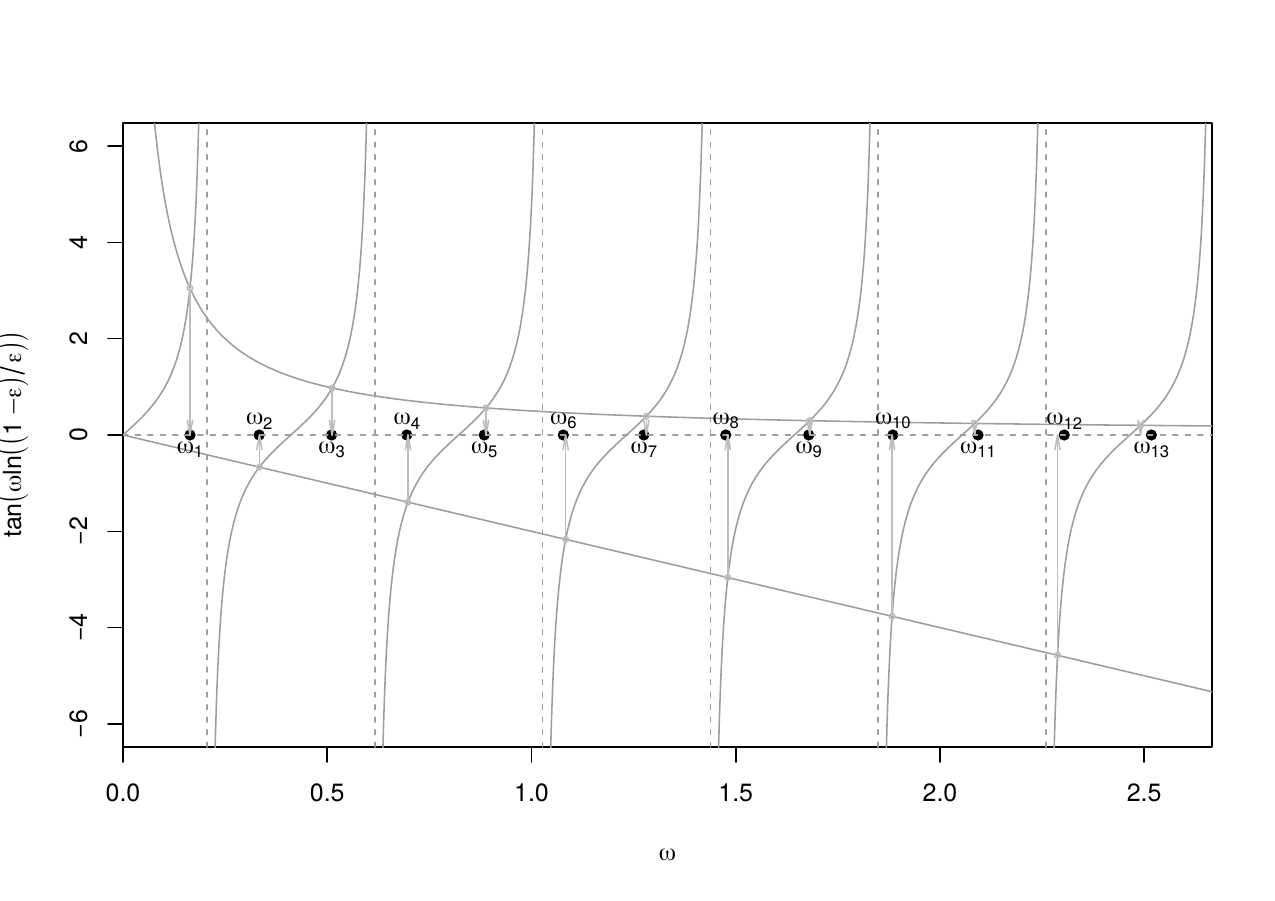}
        \caption{An illustration of asymptotic approximation.
The arrows indicate the asymptotic locations, where
the $z=\tan(\omega\ln\frac{1-\varepsilon}{\varepsilon})$
        curves intersect the curve
        $z=\frac{1}{2\omega}$ and the
        line $z=-2\omega$.
The solid dot locations are locations of $\omega$'s computed via
        eigen-decomposition. 
        }
\label{fig:eigen-01}
\end{figure}

The proof of Theorem \ref{thm:limit-dist} is given in Appendix \ref{ss:thm:limit-dist-proof}.
As shown in Figure \ref{fig:eigen-01},
the $\omega_k$'s that satisfy \eqref{eq:omega-cos}
are in the intervals
$[k\pi, k\pi+\frac{1}{2}\pi)$, one in each interval
for $k=0, 1, 2, ...$, while
the $\omega_k$'s satisfying \eqref{eq:omega-sin}
are in the intervals
$[k\pi-\frac{1}{2}\pi, k\pi)$, one in each interval
for $k=1, 2, ...$.

\begin{remark}\label{remark:Cn}
The normalizing constant $C_n$ defined in
\eqref{eq:optim-test-01}
can be reset, if desirable, by making use of
	the following bounds for $\sum_{i=1}^n\frac{1}{\lambda_i}$:
\begin{eqnarray*}
	\sum_{k=1}^n \frac{1}{\lambda_i}
\;\approx\;	\sum_{k=1}^n \frac{1}{\frac{1}{4}+
	\left[\frac{k\pi}{2\ln\frac{1-\varepsilon}{\varepsilon}}\right]^2}
\; > \; \frac{4}{\pi} \ln\frac{1-\varepsilon}{\varepsilon} \int_{\frac{\pi}{\ln\frac{1-\varepsilon}{\varepsilon}}}
	^{\frac{(n+1)\pi}{\ln\frac{1-\varepsilon}{\varepsilon}}}
	\frac{1}{1+ t^2}dt
	\;\approx\; 2\ln\frac{1-\varepsilon}{\varepsilon}
\end{eqnarray*}
and
\begin{eqnarray*}
	\sum_{k=1}^n \frac{1}{\lambda_i}
\;\approx\;	\sum_{k=1}^n \frac{1}{\frac{1}{4}+
	\left[\frac{k\pi}{2\ln\frac{1-\varepsilon}{\varepsilon}}\right]^2}
\;< \;\frac{4}{\pi} \ln\frac{1-\varepsilon}{\varepsilon} \int_0
	^{\frac{n\pi}{\ln\frac{1-\varepsilon}{\varepsilon}}}
	\frac{1}{1+ t^2}dt
\;\approx\; 2\ln\frac{1-\varepsilon}{\varepsilon}.
\end{eqnarray*}
These approximations suggest, in turn, the use of $\varepsilon\approx\frac{1}{2(n+1)}$.
\end{remark}

\begin{figure}[!htb]
\centering
\includegraphics[width=6.0in]{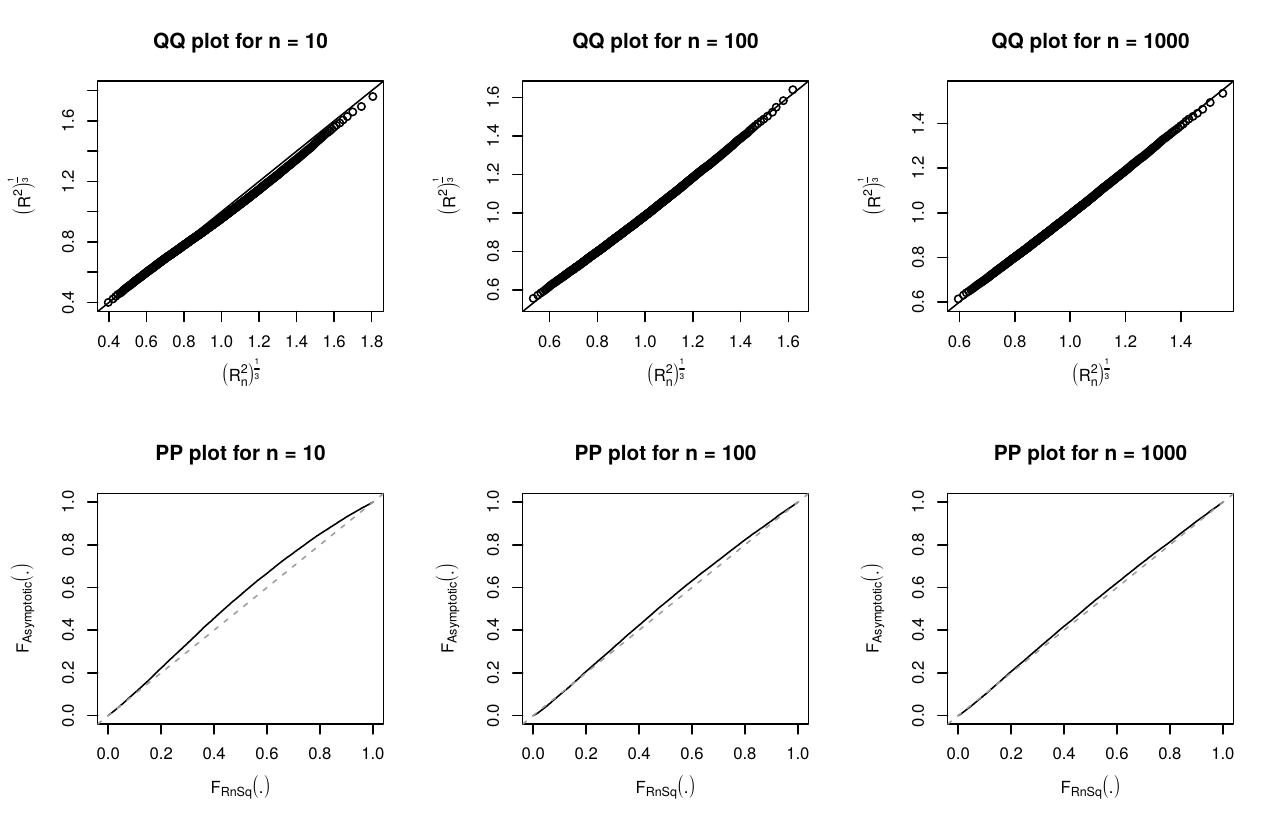}
        \caption{ Performance of asymptotic approximation
        to the distribution of $R_n^2$ for $n=10$, 100, and 1000
        obtained using a Monte Carlo approximation with 100,000 replicates.
        The three plots in the upper panel are the quantile-quantile plot
        in cubic-root scale
        with quantiles corresponding to the probabilities $i/1001$ for $i=1,...,1000$, whereas
        the three plots in the lower panel are the corresponding probability-probability plot.
        }
\label{fig:asym-eval-01}
\end{figure}

Sort all eigenvalues obtained from the $\omega_k$s in \eqref{eq:omega-cos} both 
\eqref{eq:omega-sin} 
into $\lambda_1 < \lambda_2 < ...$
Then, according to Equation (4.5) of \cite{anderson1952asymptotic},
the asymptotic distribution of $R_n^2$ is that of
\begin{equation}\label{eq:asym-dist-R_n}
	\sum_{i=1}^\infty\frac{X_i^2}{\lambda_i}
\end{equation}
where $X_i^2$s are independently and identically distributed
$\chi_1^2$ random variables.
A simple Monte Carlo simulation-based study of evaluating this asymptotic distribution
as an approximation to that of $R_n^2$ is summarized in
Figure \ref{fig:asym-eval-01}
using both quantile-quantile plots and probability-probability plots
for $n=10$, $100$, and $1,000$.
The Monte Carlo sample size used is 100,000;
The truncated series $\sum_{i=1}^n\frac{X_i^2}{\lambda_i}$ was used,
with $\varepsilon$ determined by matching $\lambda_1$.
Such $\varepsilon$  values in all three cases are close to $1/[2(n+1)]$.
It is seen from these numerical results that 
the asymptotic approximation is satisfactory even for small sample sizes.

\begin{remark}
The large-sample distribution given by \eqref{eq:asym-dist-R_n} is meant to be used for computing the critical values for significance testing in practice. Efficient computational methods are subject to future research; see \cite{davies1980algorithm} and
\cite{duchesne2010computing}. A similar remark also applies to the cases for $\tilde{W}^2_n$ 
and $\tilde{R}^2_n$, which are discussed next in Section \ref{s:asym}.
\end{remark}

\subsection{Large-Sample Results for $\tilde{W}_n^2$ and $\tilde{R}_n^2$}
\label{s:asym}
\subsubsection{Gaussian process approximation and kernel matrices}

Familiar asymptotic results for the $U_{(i)}$ process 
can be conveniently applied by writing:
\[
\sqrt{n+2}(U_{(i)}-\mu_i)\approx
\mathcal{B}(t) = W(t)-tW(1)
\mbox{ with $t=i/(n+1)$},
\]
where ``$\approx$'' means that in the limit
with $i/(n+1)\rightarrow t$, the
random variable on its left-hand side 
converges in distribution to the random variable
on its right-hand side;
See, {\it e.g.}, \cite{anderson1952asymptotic}.
We can work with this Gaussian process approximation
effectively as follows.  
Define $Z_i = \sqrt{n+1}\left[ W(i/(n+1)) - W((i-1)/(n+1))\right]$, $i=1,...,n+1$, and let $\bar{Z} = \frac{1}{n+1}\sum_{i=1}^{n+1}Z_i$.
Then $Z_i$ are {\it iid} $N(0, 1)$. 
It follows that
\begin{eqnarray*}
	&&\mathcal{B}(i/(n+1)) - \mathcal{B}((i-1)/(n+1))\\
	&=&[W(i/(n+1)) - W((i-1)/(n+1))] - [i/(n+1)-(i-1)/(n+1)]W(1)\\
	&=&[W(i/(n+1)) - W((i-1)/(n+1))] -
	\frac{1}{n+1}\sum_{j=1}^{n+1}[W(j/(n+1)) - W((j-1)/(n+1))]\\
	&=& (Z_i - \bar{Z})/\sqrt{n+1}.
\end{eqnarray*}
Let
\begin{equation}\label{eq:approx-cs-02}
	C_n = \left[\mathbf{I} -\frac{1}{n+1}\mathbf{1}\mathbf{1}'\right],
\end{equation}
a symmetric $(n+1)\times (n+1)$ matrix and the centering
operator for $Z=(Z_1,...,Z_n, Z_{n+1})'$.
Thus, for all $c=0,...,n$,
\[ 
	\sqrt{n+2}
	[U_{(i)}^{(c)}-\mu_i]
	\approx \frac{1}{\sqrt{n+1}} \tau_{c+[1:i]}'
	C_n Z
\] 
where $Z=(Z_1,...,Z_n, Z_{n+1})'\sim N_{n+1}(\mathbf{0},
\mathbf{I})$ and $\tau_{c+[1:i]}$ is the index vector for
the elements of $Z$ at $c+1, ..., c+i$, defined circularly.
Let $A_i$ be the $(n+1)\times(n+1)$ matrix obtained
by stacking these index vectors.
That is, $A_k$ is circulant with its first row consisting of
$k$ ones and $n+1-k$ zeros (in that order).
So
\[
	\sum_{c=0}^n (U_{(i)}^{(c)}-\mu_i)^2
	\approx\frac{1}{(n+1)(n+2)} Z'
	C_n
	A_i'A_i
	C_nZ
\]
and, thereby,
\[ 
	\sum_{c=0}^n \sum_{i=1}^n \psi_i
	(U_{(i)}^{(c)}-\mu_i)^2
	\approx \frac{n+1}{n+2}
	Z'
	\mathcal{K}_{n}(\psi)
	Z,
\] 
where 
\[ 
	\mathcal{K}_{n}(\psi) = \frac{1}{(n+1)^2} \sum_{i=1}^n\psi_i C_n'A_i'A_iC_n
\] 
is the $(n+1)\times (n+1)$ kernel matrix with
$\psi_i = w_i/[\mu_i(1-\mu_i)]$ for $i=1,...,n$.
To obtain the corresponding approximation to
$C(U, w)$ using the above Gaussian process,
we consider the following
Taylor expansion of the $i$-th summand of Eq. \eqref{eq:normalized-Wn} 
at $\mu_i$:
\[
	-2\left[\mu_i\ln\frac{U_{(i)}}{\mu_i}
+(1-\mu_i)\ln\frac{1-U_{(i)}}{1-\mu_i} \right]
\approx \frac{1}{\mu_i(1-\mu_i)} \left(U_{(i)}-\mu_i\right)^2.
	\]
This suggests the following approximation to the 
	circularly symmetric test statistic $C(U, w)$:
\[
	C(U, w) \approx \frac{1}{n+2}
	Z'\mathcal{K}_{n}(\psi)Z
\]
or, simply, $C(U, w) \approx \frac{1}{n}Z'\mathcal{K}_{n}(\psi)Z.$

Following \cite{anderson1952asymptotic}, we denote by
$1/\lambda_i$ the eigenvalues of
$\frac{1}{n+2}\mathcal{K}_{n}(\psi)$ 
and the corresponding eigenvectors $V_i$.
That is, we can write
\[ 
	C(U, w)
	\approx \sum_{i=1}^{n+1}\frac{1}{\lambda_i} Z'V_iV_i'Z
\] 
a weighted sum of independent $\chi_1^2$'s with weights
$1/\lambda_i$'s;
See \cite{anderson1952asymptotic},
\cite{stephens1974edf},
\cite{sinclair1988approximations},
\cite{zolotarev1961concerning},
\cite{davies1980algorithm},
and \cite{duchesne2010computing}
for
the case of its continuum limit, and
numerical methods if desirable.

Computationally, the large-sample approximation relies on
the eigenvalue decomposition of the kernel matrix
$\mathcal{K}_{n}(\psi)$.
The easy-to-prove results summarized into the following 
proposition show that the kernel matrix  $\mathcal{K}_{n}(\psi)$
is circulant \citep[see, {\it e.g.},][]{CIT-006}.

\begin{prop}
Consider the $(n+1)\times (n+1)$ matrix $C_n$ defined in \eqref{eq:approx-cs-02}
and the $(n+1)\times (n+1)$
	matrices $A_k$, $k=1,...,n$, defined above.
	Then
\begin{enumerate}
\item[\rm (a)] $C_n$ and $A_k$'s are all circulant matrices, so are
	$A_k'A_k$ and $C_n'A_k'A_kC_n$;
\item[\rm (b)] The matrix
$A_k'A_k$ is a symmetric Toeplitz matrix, with
the $(i,j)$'s elements:
\[\max(0, k+i-j) +\max(0, k-i+j-(n+1))
\] for $1\leq i\leq j\leq n+1$; 
\item[\rm (c)] for all $k=1,...,n$,
\[
	C_n'A_k'A_k C_n
	= A_k'A_k -\frac{k^2}{n+1}\mathbf{1}\mathbf{1}';
\]
and
\item[\rm (d)] the kernel matrix $\mathcal{K}_{n}(\psi)$ is symmetric and
	circulant with elements: 
\begin{equation}\label{eq:CS-cov-01}
	\frac{1}{(n+1)^2}\sum_{k=1}^n
	\psi_k
	\left[\max(0, k+i-j) +\max(0, k-i+j-(n+1))
	-\frac{k^2}{n+1}\right]
\end{equation}
for $1\leq i\leq j\leq n+1$.
\end{enumerate}
\end{prop}




The following theorem summarizes the properties
of the eigenvalues and eigenvectors of circulant matrices
\citep[Theorem 7 of][]{CIT-006}.
\begin{thm}
Let $C$ be a $(n+1)\times (n+1)$ circulant matrix
with its first row denoted by $c=(c_0,...,c_n)'$,
	{\it i.e.}, the $(k,j)$ entry of
	$C$ is given by $C_{k,j}=c_{(j-k)\mbox{ mod } (n+1)}$.
Then $C$ has the eigenvectors
\[
	v^{(m)} = \frac{1}{\sqrt{n+1}}
	(1, e^{-2\pi i m/(n+1)}, e^{-2\pi i 2m/(n+1)},
	\ldots, e^{-2\pi i nm/(n+1)})' , \quad m=0,1,\ldots,n,
\]
and corresponding eigenvalues
\[
	\phi_m = \sum_{k=0}^n c_k e^{-2\pi i k m/(n+1)}
\]
and can be expressed in the form $C=V\Phi V^*$, 
where $i$ is the unit imaginary number,
$V$ has the eigenvectors as columns in order,
the asterisk * denotes conjugate transpose, 
and 
$\Phi$ is $\mbox{diag}(\phi_k)$.  In particular, all circulant
matrices share the same eigenvectors, the same matrix
$U$ works for all
circulant matrices, and any matrix of the form 
$C=V\Phi V^*$ is circulant.
Furthermore, for any two $(n+1)\times (n+1)$ circulant matrices
	$C$ and $B$, $C$ and $B$ commute, {\it i.e.},
	$CB=BC$, and $CB$, $\alpha C$, and $C+B$ are circulant matrices,
	where $\alpha$ is a scalar.
\end{thm}

Since the eigenvalues of any real symmetric matrix are real,
the symmetric circulant matrix $\mathcal{K}_{n}(\psi)$
has $(n+1)$ real 
eigenvalues
\[ 
	\phi_m = \sum_{k=0}^n c_k \cos\left(\frac{2\pi k m}{n+1}\right).
\] 
Moreover, a $(n+1)\times (n+1)$ symmetric circulant matrix $C$ satisfies the extra condition
that $c_{n-i}=c_{i}$
and is thus determined by $\lfloor(n+1)/2\rfloor+1$ elements.
The corresponding eigenvalues can be written as
\[
	\phi_m = c_0+2\sum_{k=1}^{(n+1)/2-1} c_k
	\cos\left(\frac{2\pi k m}{n+1}\right)
	+ c_{(n+1)/2} \cos\left(\pi k m\right)
	\]
for $(n+1)$ even, and
\[
	\phi_m = c_0+2\sum_{k=1}^{n/2} c_k
	\cos\left(\frac{2\pi k m}{n+1}\right)
	\]
for $(n+1)$ odd.
These properties allow for efficient computation
via fast discrete Fourier transform
\citep{cooley1965algorithm}.
The large-sample approximation to
\eqref{eq:CS-cov-01} for $\tilde{W}_n^2$ and 
$\tilde{R}_n^2$ is given in the next two subsections.

\subsubsection{Circularized $W_n^2$, $\tilde{W}_n^2$}
\label{ss:Wn-asym}
The continuum limit of the kernel structure
\eqref{eq:CS-cov-01} for the test statistic $W_n^2$
can be obtained and is summarized into the kernel function
in the following theorem;
See Appendix \ref{proof:Wn} for the proof.

\begin{thm}\label{thm:AD-limit-kernel}
	If $w_k=1$, {\it i.e.},  $\psi_k \propto 1/[\mu_k(1-\mu_k)]$,
then in its continuum limit the kernel matrix
$\mathcal{K}_{n}(\psi)$
	with elements \eqref{eq:CS-cov-01}
is given by
\begin{eqnarray}
	\kappa(t,s)
	&=&
	\left\{\begin{array}{lll}
	2\left[(s-t) \ln(s-t) +[1 - (s-t)] \ln(1-(s-t))\right]+1,
		&&\mbox{if $ t\leq s$};\\
	2\left[(t-s) \ln(t-s) +[1 - (t-s)] \ln(1-(t-s))\right]+1,
		&&\mbox{if $ s < t$}.
	\end{array}\right.
	\label{eq:Wn-asym-kernel}
\end{eqnarray}
\end{thm}
\newboolean{longversion}
\newboolean{shortversion}
\setboolean{shortversion}{true}

\begin{remark}\label{remark-asym-01}
In the case with $w_k=1$, {\it i.e.}, $\psi_k = 1/[\mu_k(1-\mu_k)]$,
the diagonal elements of the kernel matrix are given
by
\[ 
	\sum_{k=1}^n \frac{k-\frac{k^2}{n+1}}{k(n+1-k)}
	= \frac{n}{n+1}.
\] 
This implies that the sum of all the eigenvalues of the kernel matrix is $n$.
Because the rank of the circulant matrix $C_n$ is $n$ and 
	the circulant matrix $\sum_{k=1}^n\psi_kA_k'A_k$ is full rank,
the kernel matrix is a rank-$n$ matrix and thus has 
$n$ nonzero eigenvalues with zero eigenvalue given by
\[
	\phi_0 = \sum_{k=0}^n c_k,
\]
where $c=(c_0,...,c_n)$ denotes the first row of the kernel matrix.
\end{remark}

\begin{figure}[!htb]
\centering
\includegraphics[width=6.0in]{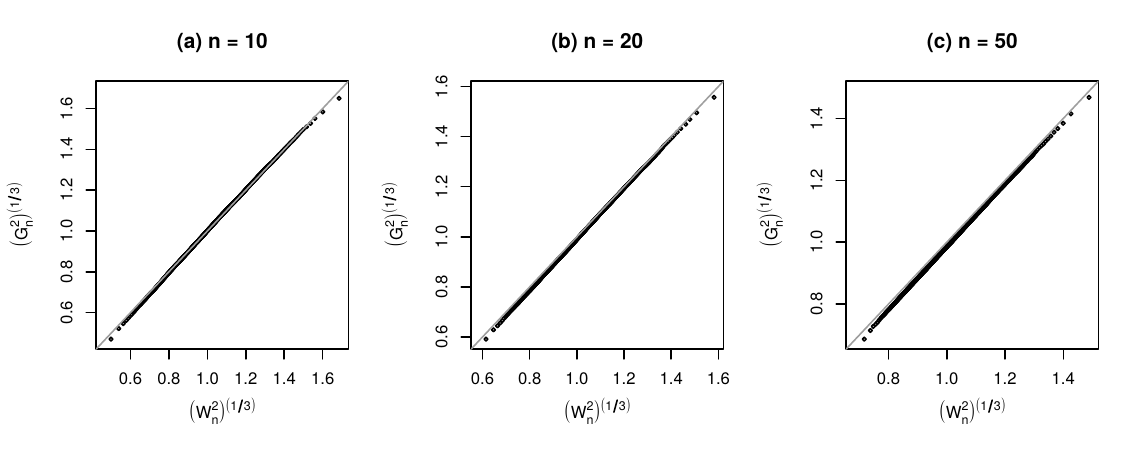}
        \caption{Quantile-Quantile plot of
the large-sample approximation to the distribution of $\tilde{W}_n^2$
        versus the true distribution in the cubic-root scale,
        obtained on 1,000,000 Monte Carlo samples.
The quantile points are obtained for
1,000 equally spaced probabilities from $1/1001$ to $1-1/1001$.
        }
\label{fig:asym-01}
\end{figure}

Remark \ref{remark-asym-01}
implies that
$E(C(U,w))\approx 1$
and, naturally, suggests
$\tilde{W}_n^2$ defined in \eqref{eq:normalized-Wn}
for the preference of $E(\tilde{W}_n^2)\approx 1$.
A numerical evaluation of the large-sample-based approximation
to the distribution of $\tilde{W}_n^2$ 
is shown by the quantile-quantile plots in Figure \ref{fig:asym-01} for
a selected cases of $n=10$, $20,$ and $50$.
The quantile points are obtained for
1,000 equally spaced CDF values from  $1/1001$ to $1-1/1001$
based on a Monte Carlo approximation of 1,000,000 replicates.
The asymptotic kernel function \eqref{eq:Wn-asym-kernel}
is used to compute $n+1$ values as the first row of 
a corresponding kernel matrix:
\[
	c_0 = \frac{1}{n+1},
\quad c_k = \frac{2\left[\frac{k}{n+1}\ln\frac{k}{n+1}
+\frac{n+1-k}{n+1}\ln\frac{n+1-k}{n+1}\right] + 1}{n+1},\quad
\quad k=1,...,n.
	\]
One may see from Figure \ref{fig:asym-01} that the approximation is satisfactory.

\subsubsection{Circularized $R_n^2$, $\tilde{R}_n^2$}
The continuum limit of the kernel structure
\eqref{eq:CS-cov-01} for the test statistic $R_n^2$
can also be obtained and is summarized into the kernel function
in the following theorem; See Appendix \ref{proof:Rn} for the proof.


\begin{thm}\label{thm:New-limit-kernel}
	If $w_k=1/[\mu_k(1-\mu_k)]$, {\it i.e.},
	$\psi_k \propto 1/[\mu_k(1-\mu_k)]^2$,
then in its continuum limit the kernel matrix
$\mathcal{K}_{n}(\psi)$
	with elements \eqref{eq:CS-cov-01}
is given by
\begin{eqnarray}
	\kappa(t,s)
&=&
	\left\{\begin{array}{lll}
	2\left[2(s-t)-1\right] \ln\frac{s-t}{1-(s-t)} - 2,
		&&\mbox{if $ t\leq s$};\\
	2\left[2(t-s)-1\right] \ln\frac{t-s}{1-(t-s)} - 2,
		&&\mbox{if $ s < t$}.
	\end{array}\right.
	\label{eq:Rn-asymp-kernal}
\end{eqnarray}
\end{thm}

\begin{remark}\label{remark-asym-02}
In case with $w_k=1/[\mu_k(1-\mu_k)]$, {\it i.e.},
$\psi_k = 1/[\mu_k(1-\mu_k)]^2$,
the diagonal elements of the kernel matrix are given
by
\[ 
\frac{1}{(n+1)^2}\sum_{k=1}^n \psi_k C_n'A_k'A_kC_n\\
	=(n+1)^2\sum_{k=1}^n \frac{k-\frac{k^2}{n+1}}{[k(n+1-k)]^2}
	=2\sum_{k=1}^n\frac{1}{k}
	\] 
This implies that the sum of all the eigenvalues of the kernel matrix is
	is $2(n+1)\sum_{k=1}^n\frac{1}{k}$.
\end{remark}

\begin{figure}[!htb]
\centering
\includegraphics[width=6.0in]{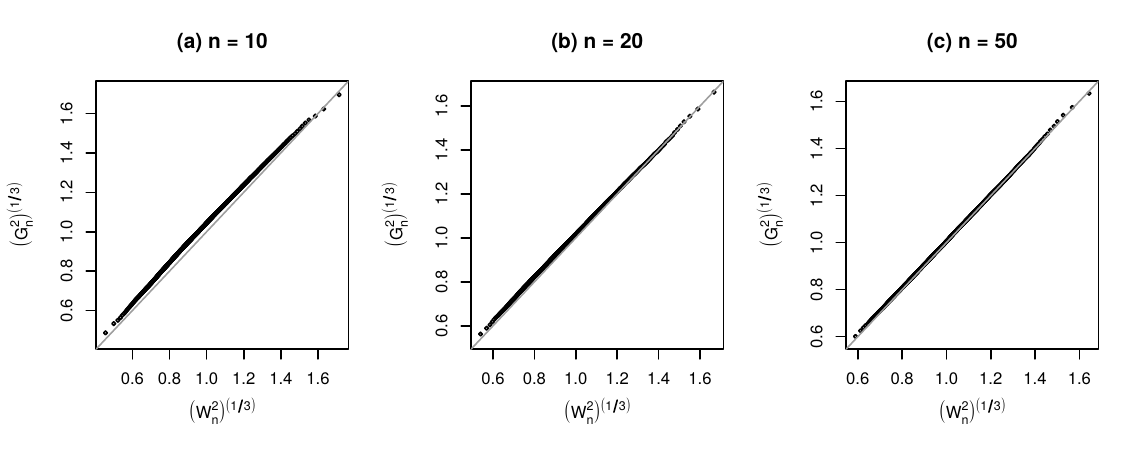}
        \caption{
                The legend is the same as that of \ref{fig:asym-01},
                except that the test statistic is $\tilde{R}_n^2$.
        }
\label{fig:asym-02}
\end{figure}

Remark \ref{remark-asym-02}
implies that
$E(B(U,w)) = 2\sum_{k=1}^n\frac{1}{k}$
and, naturally, suggests
the test statistic $\tilde{R}_n^2$ defined in
\eqref{eq:normalized-Rn}
for the preference of $E(\tilde{R}_n^2)\approx 1$.
A numerical evaluation of the large sample-based approximation
to the distribution of $\tilde{R}_n^2$
was conducted in the same way as that for $\tilde{W}_n^2$
in the previous subsection. 
The asymptotic kernel function \eqref{eq:Rn-asymp-kernal}
is used to compute $n+1$ values as the first row of 
a corresponding kernel matrix:
\[
	c_0 = \frac{1}{2(n+1)\sum_{k=1}^n\frac{1}{k}},
\quad c_k = \frac{\left(\frac{2k}{n+1}-1\right)\ln\frac{k}{n+1-k}-1}{(n+1)\sum_{k=1}^n\frac{1}{k}},
\quad k=1,...,n.
	\]
The results are displayed in Figure \ref{fig:asym-02}, which
shows that the approximation is satisfactory, although this can be
further improved by small sample corrections.



\section{Concluding Remarks}

\label{s:conclusion}


Assessing goodness-of-fit is a fundamental problem in both applied and theoretical
statistics in general, and in data-driven (or auto-)modeling in
contemporary big data analysis in particular.
This paper aimed to three goals toward both  deep understanding
of the problem and perfection of the Anderson-Darling test. 
It provided a geometric intuition for understanding,
which leads to the conclusion that $R_n^2$ can serve
as an omnibus
test. This is consistent with the discovery of \cite{zhang2002powerful}.
Furthermore, it proposed the method of circularization
and showed that circularized versions can have a better performance
than their parent tests.
In addition, this paper also established the asymptotic distributions
of $R_n^2$ and the two circularly symmetric tests
$\tilde{W}_n^2$ and $\tilde{R}_n^2$, although more theoretical investigations are needed.

Performance of the proposed methods can also be investigated for
distributions containing unknown parameters. This can be 
done with either the traditional approach, which replies on
point estimations of the unknown parameter, or
the inferential models approach of \cite{martin2015inferential},
which can be viewed as a generalized theory of 
the familiar method of pivotal quantity
for constructing confidence intervals and 
hypothesis testing.

\ifthenelse{1=0}{}{

\section*{Acknowledgements}

The author would like to express his sincere gratitude to the Editor, Associate Editor, and two reviewers for their thorough and thoughtful comments, which have greatly improved the quality and clarity of this revised manuscript. He would also like to thank Professor Yaowu Liu for his helpful and valuable comments on an early version of the manuscript. Their insightful and constructive comments have been instrumental in shaping and refining this article.

}

\bibliography{omnibus}

\appendix

\section{Proofs of Theorems}

\ifthenelse{1=1}{}{
\subsection{Proof of Lemma  \ref{lemma:weights-and-direction}}
\label{app:weights-and-direction}
The proof is straightforward by applying the Cauchy–Schwarz inequality
to the inner product of
$\Sigma^{\frac{1}{2}}w$ and $\Sigma^{-\frac{1}{2}}\delta$:
\[
\frac{(w'\delta)^2}{w'\Sigma w}
=\frac{[(\Sigma^{\frac{1}{2}}w)'(\Sigma^{-\frac{1}{2}}\delta)]^2}{w'\Sigma w}
\leq 
\frac{(w'\Sigma w)(\delta\Sigma^{-1}\delta)}{w'\Sigma w}
=\delta\Sigma^{-1}\delta,
\]
where the equality holds
if and only if
$\Sigma^{\frac{1}{2}}w \propto \Sigma^{-\frac{1}{2}}\delta$,
that is,
$w \propto \Sigma^{-1}\delta.$ 
}

\subsection{Proof of Theorem \ref{thm:exact-cov}}
\label{ss:proof-exact-cov}

Using the popular technique for deriving the expectation of $\ln(X)$ when $X$ is a Beta random variable, we have
\[
	\mbox{E}[\ln(U_{(i)})] 
=\left. \frac{1}{\mbox{Beta}(\alpha, \beta)}\frac{\partial}{\partial \alpha}
\int_0^1u^{\alpha-1}(1-u)^{\beta-1}du
\right|_{\alpha=i, \beta=n+1-i}
= \psi(i) - \psi(n+1).
	\]
The claimed results can be verified using such standard techniques
with tedious algebraic operations.

\subsection{Proof of Theorem \ref{thm:finite-cov}}
\label{app:cov}

From the pdf of the
	joint distribution $U_{(i)}$ and $U_{(j)}$,
	we have for all $k$ and $l$,
	\[
              E\left\{U_{(i)}^k[1-U_{(j)}]^\ell \right\}
        = \frac{ (i+k-1)!  }{(i-1)!} \frac{(n-j+\ell)!}{(n-j)!} \frac{ n!  } {(n+k+\ell)!}.
\]
With this identify and tedious routine algebraic operations,
one can verify the claimed results.

\subsection{Proof of Theorem \ref{thm:limit-cov}}
\label{app:limit-cov}

It is known \citep[see, {\it e.g.},]{shorack1972functions} that as $n\rightarrow\infty$, $\mathcal{B}_n(t)$
converges in distribution to a Brownian bridge. 
The results on the covariance structure of the Brownian bridge
are well-known and easy-to-prove. So, our proof here will focus on the
results on $\mbox{Cov}\left(\mathcal{B}^2(s), \mathcal{B}^2(t)\right)$.

Write the Brownian bridge using the Brownian motion $W(t)$,
$t \in [0, 1]$ as follows
	\[
\mathcal{B}(t) = W(t) - tW(1).
		\]
Thus
\[
E(\mathcal{B}^2(t))
	= E([(1-t)W(t) - t(W(1)-W(t))]^2)
	= t(1-t)
\]
For $0 <s < t < 1$, it is easy to see that
\begin{eqnarray*}
	E\left[W^2(s)W^2(t)\right]
	= 3s^2 + s(t-s),
\end{eqnarray*}
and
\begin{eqnarray*}
	E\left[\mathcal{B}^2(s)\mathcal{B}^2(t)\right]
	= 2s^2+st - 5s^2t-st^2 + 3s^2t^2.
\end{eqnarray*}
Thus, the result follows.

\subsection{Proof of Theorem \ref{thm:aymptotic-weights}}
\label{app:symp-optim}

The variance of the Taylor expansion \eqref{eq:RnW-approx-03}, {\it i.e.}, the variance of
\[
r_n^2(w) = \sum_{i=1}^{n}\replace{\frac{w_i}{\mu_i(1-\mu_i)}}{w_i}(U_{(i)}-\mu_i)^2,
\]
can be written as
\[
\mbox{Var}(r_n^2(w))=\sum_{i=1}^{n}\sum_{j=1}^{n}\replace{\frac{w_i}{\mu_i(1-\mu_i)}}{w_i}\mbox{Cov}\left((U_{(i)}-\mu_i)^2,
(U_{(j)}-\mu_j)^2\right)\replace{\frac{w_j}{\mu_j(1-\mu_j)}}{w_j}.
\]
Applying the method of Lagrange multipliers, we see that
the optimal weight vector $w^{\mbox{(optimal)}}$ is given by
\[
\sum_{j=1}^n w_j\mu_j(1-\mu_j)=1\quad \mbox{ and }\quad 
\sum_{j=1}^{n}\mbox{Cov}\left((U_{(i)}-\mu_i)^2,(U_{(j)}-\mu_j)^2\right)w_j = \lambda
\replace{}{\mu_i(1-\mu_i)}
\]
for some $\lambda$ and all $i=1,...,n.$ 
Making use of Theorem \ref{thm:limit-cov}, we have the corresponding continuum limit 
\begin{eqnarray*}
&&2\int_0^st^2(1-s)^2\psi(t)dt+2\int_s^1s^2(1-t)^2\psi(t)dt \\
&=&2(1-s)^2\int_0^st^2\psi(t)dt+2s^2\int_s^1(1-t)^2\psi(t)dt \\
&=& \replace{\lambda}{4\lambda s(1-s)}
\end{eqnarray*}
for $s \in [\varepsilon, 1-\varepsilon]$ and $\varepsilon\in (0, 1/2)$.
Thus, we differentiate both sides of the above equation with respect to $s$ to obtain the continuum limit of the Lagrange auxiliary equation for $\psi(t)$:
\begin{equation}\label{eq:lag-02}
	-4(1-s)\int_{0}^s t^2 \psi(t)dt
	+4s\int_s^{1} (1-t)^2 \psi(t)dt
	= \replace{0}{4\lambda(1-2s)}
\end{equation}
because
\(
	2(1-s)^2 s^2 \psi(s) -2s^2 (1-s)^2 \psi(s) = 0
\)
for all $s \in [\varepsilon, 1-\varepsilon]$. 
Equation \eqref{eq:lag-02} 
is an integral equation, known as the Fredholm equation, which does not
have a general solution. Here, we solve it by converting it into a 
differential equation.

Differentiate both sides of Equation \eqref{eq:lag-02} with respect to $s$ to obtain
\begin{equation}\label{eq:lag-102}
	\int_{0}^s t^2 \psi(t)dt
	+\int_s^{1} (1-t)^2 \psi(t)dt
	-s(1-s) \psi(s) = \replace{0}{-8\lambda}.
\end{equation}
Differentiating the two sides of Equation \eqref{eq:lag-102} with respect to $s$, we obtain
\begin{equation}\label{eq:lag-103}
	\psi'(s) -2\left[\frac{1}{1-s}-\frac{1}{s}\right] \psi(s) = 0.
\end{equation}
Applying the method of separation of variables,
we get the solution to Equation \eqref{eq:lag-103}:

\ifthenelse{1=1}{}{
\color{red}

We show that the solution $w(t)$ to \eqref{eq:lag-02} is symmetric about
$t=\frac{1}{2}$.
Let $g(t)=w(1-t)$. Then $w(t) = g(1-t)$ and,
with the change-variable method through transformation $u=1-t$,
\begin{eqnarray*}
	&&-4(1-s)\int_{0}^s t^2 w(t)dt
	+4s\int_s^{1} (1-t)^2 w(t)dt
	\\
&=&	-4(1-s)\int_{0}^s t^2 g(1-t)dt
	+4s\int_s^{1} (1-t)^2 g(1-t)dt
	\\
&=&	-4(1-s)\int_{1-s}^{1} (1-u)^2 g(u)du
	+4s\int_{0}^{1-s} u^2 g(u)du
	\\
&=&	-4s'\int_{s'}^{1} (1-u)^2 g(u)du
	+4(1-s')\int_{0}^{s'} u^2 g(u)du
\end{eqnarray*}
for all $s'=1-s \in [\varepsilon, 1-\varepsilon]$.
The last identity implies that $w(t)$ is symmetric about
$t=\frac{1}{2}$, assuming that the solution to \eqref{eq:lag-02} exits and unique.
In this case, we only need to consider the case of $s\in [\varepsilon, 1/2]$.
Let $s= \frac{1}{2}-\delta$, $\delta \in [0, 1/2-\varepsilon]$.
Write \eqref{eq:lag-02} as follows
\begin{eqnarray*}
	&&-4\left(\frac{1}{2}+\delta\right)
	\int_{0}^{\frac{1}{2}-\delta} t^2 w(t)dt
	+4\left(\frac{1}{2}-\delta\right) \int_{\frac{1}{2}-\delta}^{1} (1-t)^2 w(t)dt
	\\
&=&2\left[ \int_{\frac{1}{2}-\delta}^{1} (1-t)^2 w(t)dt
	-\int_{0}^{\frac{1}{2}-\delta} t^2 w(t)dt \right]
	-4\delta\left[ \int_{0}^{\frac{1}{2}-\delta} t^2 w(t)dt
	+ \int_{\frac{1}{2}-\delta}^{1} (1-t)^2 w(t)dt \right]
	\\
&=&2 \int_{\frac{1}{2}-\delta}^{\frac{1}{2} + \delta} (1-t)^2 w(t)dt
	-4\delta\left[
	2	\int_{0}^{\frac{1}{2}} t^2 w(t)dt
		-\int_{\frac{1}{2}-\delta}^{\frac{1}{2}} t^2 w(t)dt
	+ \int_{\frac{1}{2}-\delta}^{\frac{1}{2}} (1-t)^2 w(t)dt
	\right]
	\\
&=&2 \int_{\frac{1}{2}}^{\frac{1}{2} + \delta} t^2 w(t)dt
	+ 2 \int_{\frac{1}{2}}^{\frac{1}{2} + \delta} (1-t)^2 w(t)dt
\\
	&&
	-4\delta\left[
		2	\int_{\frac{1}{2}}^{1} (1-t)^2 w(t)dt
		-\int_{\frac{1}{2}}^{\frac{1}{2}+\delta} (1-t)^2 w(t)dt
	+ \int_{\frac{1}{2}}^{\frac{1}{2}+\delta} t^2 w(t)dt
	\right]
	\\
	&=& 2 \int_{0}^{\delta} \left(\frac{1}{2}+x\right)^2 g(x)dx
	+ 2 \int_{0}^{\delta} \left(\frac{1}{2}-x\right)^2 g(x)dx
\\
	&&
	-4\delta\left[
		2	\int_0^{\frac{1}{2}} \left(\frac{1}{2}-x\right)^2 g(x)dx
		-\int_{0}^{\delta} \left(\frac{1}{2}-x\right)^2 g(x)dx
	+ \int_{0}^{\delta} \left(\frac{1}{2}+x\right)^2 g(x)dx
	\right]
	\\ &=& 0,
\end{eqnarray*}
where $g(x) = w(x+1/2)$, $x \in [0, 1/2-\varepsilon]$.
\begin{eqnarray*}
	&& 2 \int_{0}^{\delta} \left(\frac{1}{2}+x\right)^2 g(x)dx
	+ 2 \int_{0}^{\delta} \left(\frac{1}{2}-x\right)^2 g(x)dx
\\
	&&
	-4\delta\left[
		2	\int_0^{\frac{1}{2}} \left(\frac{1}{2}-x\right)^2 g(x)dx
		-\int_{0}^{\delta} \left(\frac{1}{2}-x\right)^2 g(x)dx
	+ \int_{0}^{\delta} \left(\frac{1}{2}+x\right)^2 g(x)dx
	\right]
	\\
&=&  \int_{0}^{\delta} \left(1+4x^2\right) g(x)dx
	-4\delta\left[
		2	\int_0^{\frac{1}{2}} \left(\frac{1}{2}-x\right)^2 g(x)dx
	+ 2 \int_{0}^{\delta} x g(x)dx
	\right]
	\\
&=&  \int_{0}^{\delta} \left(1+4x^2\right) g(x)dx
	-8\delta \int_0^{\frac{1}{2}} \left(\frac{1}{2}-x\right)^2 g(x)dx
	- 8\delta \int_{0}^{\delta} x g(x)dx
	\\
	&=& 0.
\end{eqnarray*}
Differentiating the two sides of the above equation gives the result
\begin{eqnarray*}
&&	(1+4\delta^2) g(\delta) -8\int_0^{\frac{1}{2}} \left(\frac{1}{2}-x\right)^2 g(x)dx -8\delta^2 g(\delta)
	- 8 \int_{0}^{\delta} x g(x)dx
	\\
	&=&	4\left(\frac{1}{2}-\delta\right)
		\left(\frac{1}{2}+\delta\right) g(\delta) -8
		\int_0^{\frac{1}{2}} \left(\frac{1}{2}-x\right)^2 g(x)dx
	- 8 \int_{0}^{\delta} x g(x)dx
	\\
	&=& 0.
\end{eqnarray*}

Let 
\[ \psi(\delta) = \left(\frac{1}{2}-\delta\right) \left(\frac{1}{2}+\delta\right) g(\delta)
	\]
	{\it i.e.}, write
\[ g(x) = \frac{\psi(x)}{\left(\frac{1}{2}-x\right) \left(\frac{1}{2}+x\right)}.
	\]
We get
\begin{eqnarray*}
	&&	4\left(\frac{1}{2}-\delta\right)
		\left(\frac{1}{2}+\delta\right) g(\delta) 
		-8 \int_0^{\frac{1}{2}} \frac{\left(\frac{1}{2}-x\right)^2}
        {\left(\frac{1}{2}-x\right) \left(\frac{1}{2}+x\right)} \psi(x)dx
	- 8 \int_{0}^{\delta} x g(x)dx
	\\
&=& 4\psi(\delta) - 8
	\int_0^{\frac{1}{2}} \frac{\left(\frac{1}{2}-x\right)^2}
	{\left(\frac{1}{2}-x\right) \left(\frac{1}{2}+x\right)} \psi(x)dx
	-8\int_{0}^{\delta}
	\frac{x}{\left(\frac{1}{2}-x\right)\left(\frac{1}{2}+x\right)}
	\psi(x)dx
	\\
	&=& 0
\end{eqnarray*}
for all $\delta \in [0, 1/2-\varepsilon]$.
Differentiating the two sides of the above equation, we have
\begin{equation}\label{eq:lag-08}
	\psi'(x)  - 
	\frac{2x}{\left(\frac{1}{2}-x\right)\left(\frac{1}{2}+x\right)}
	\psi(x) = 0
	\qquad(x\in [0, 1/2).
\end{equation}
Applying the method of separation of variables,
rearranging terms in \eqref{eq:lag-08} to obtain the equation
\[
	\frac{d\psi}{\psi}
	= \frac{2x}{\left(\frac{1}{2}-x\right)\left(\frac{1}{2}+x\right)}dx
	\]
and integrating  
the both sides of this equation
\[
	\int \frac{d\psi}{\psi}
= \int \frac{2x}{\left(\frac{1}{2}-x\right)\left(\frac{1}{2}+x\right)}dx,
	\]
we get the solution to \eqref{eq:lag-08}
\begin{equation}\label{eq:lag-18}
	\psi(x) = \frac{c_0}{\left(\frac{1}{2}-x\right)\left(\frac{1}{2}+x\right)}
\end{equation}
for some positive constant $c_0$.
Thus
\begin{equation}\label{eq:lag-18}
	w(t) = g\left(t-\frac{1}{2}\right)
	=\frac{1}{t\left(1-t\right)} \psi\left(t-\frac{1}{2}\right)
	= \frac{1}{t\left(1-t\right)} \frac{c_0}{t\left(1-t\right)}
	= \frac{c_0}{t^2\left(1-t\right)^2}
\end{equation}
for $t \in [1/2, 1-\varepsilon]$.
Making use of the property that $w(.)$ is symmetric
about $t=1/2$, we have from \eqref{eq:lag-18}
\color{black}
}
\begin{equation}\label{eq:lag-20}
	\psi(t) = \frac{c_0}{t^2\left(1-t\right)^2},
	\qquad (t \in [\varepsilon, 1-\varepsilon])
\end{equation}
for some positive constant $c_0$.

Note that $\psi(.)$ on $(0,\varepsilon)$ and
$(1-\varepsilon, 1)$ must satisfy condition \eqref{eq:lag-02}, that is,
\begin{equation}\label{eq:lag-28}
	-4(1-s)\int_0^{\varepsilon} t^2 \psi(t)dt
	+4s\int_{1-\varepsilon}^1 (1-t)^2 \psi(t)dt
	-4(1-s)\int_{\varepsilon}^s \frac{c_0}{(1-t)^2} dt
	+4s\int_s^{1-\varepsilon} \frac{c_0}{t^2}dt
	= 0
\end{equation}
for all $s\in [\varepsilon, 1-\varepsilon]$.
We need to show that such a $\psi(.)$ exists.
Taking $\psi(t) = \frac{c_0 h(\varepsilon)}{t^2(1-t)^2}$ with
$h(\varepsilon)=1/\varepsilon$, for example, we have
for the left-hand side of \eqref{eq:lag-28}:
\begin{eqnarray*}
&& 4[s -(1-s)]\int_0^{\varepsilon} \frac{c_0 h(\varepsilon)}{(1-t)^2}dt
	-4(1-s)\left[\left.\frac{c_0}{1-t}\right|_{\varepsilon}^s\right]
	+4s\left[\left.-\frac{c_0}{t}\right|_{s}^{1-\varepsilon}\right]
	\\
&=& 4(2s -1)c_0h(\varepsilon)\frac{\varepsilon}{1-\varepsilon}
	+4(1-2s) c_0 \frac{1}{1-\varepsilon}
	\\
	&=& 0.
\end{eqnarray*} 
Now, letting $\varepsilon\rightarrow 0$, we obtain
from \eqref{eq:lag-20} that
\[ 
	\psi(t) = \frac{c_0}{t^2\left(1-t\right)^2}
	\qquad(t \in (0, 1)).
\] 
That is, the solution $\psi(t)$ to Equation \eqref{eq:lag-02} is proportional to $\frac{1}{t^2\left(1-t\right)^2}$, the same as \eqref{eq:asym-optim-weights}. This completes the proof.


\subsection{Proof of Theorem \ref{thm:limit-dist}}
\label{ss:thm:limit-dist-proof}

Let $h(t) = f(t)\psi^{-1/2}(t)$.
Recall from \cite{anderson1952asymptotic} that
\[h''(t) + \lambda \psi(t) h(t)=0. \]
In the proof, we work with the following transformation
\[ x = 2t-1\quad \mbox{ and, thereby, } \quad t = (1+x)/2,
\qquad x \in [-1+2\varepsilon, 1-2\varepsilon].
	\]
So we have
$\psi(t=(1+x)/2) = 16/(1-x^2)^2$ and
$g(x) = f(t=(1+x)/2) \psi^{-1/2}(t=(1+x)/2)
= \frac{1}{4}(1-x^2)f(t=(1+x)/2)
= \frac{1}{4}(1-x^2) \tilde{f}(x).$
Thus,
\[g'(x)  = \frac{1}{4}\left[ (1-x^2) \tilde{f}'(x) -2x\tilde{f}(x)\right]
\]
and
\[g''(x)  = \frac{1}{4}\left[
	(1-x^2) \tilde{f}''(x) -4x \tilde{f}'(x) -2\tilde{f}(x)
	\right].
\]
It follows from
\begin{equation}\label{eq:Sturm-Liouville-02}
	g''(x) + \frac{4\lambda}{(1-x^2)^2}  g(x) = 0 
\end{equation}
that
\begin{equation}\label{eq:R_nSqDE-01}
 (1-x^2) \tilde{f}''(x) -4x \tilde{f}'(x) 
	 +	\left[\frac{4\lambda}{1-x^2}-2\right]  \tilde{f}(x) = 0 
\end{equation}


The second-order differential equation \eqref{eq:R_nSqDE-01}
can be solved with Mathematica \citep[][]{Mathematica}
or the trial solution method with
\[ 
 y(x) = c (1-x^2)^{-\tau}e^{\xi\arctanh(x)}
	\]
where $c$, $\tau$, and $\xi$ are constant, $\atanh(.)$ is the inverse of the hyperbolic function 
$\tanh$: \[\atanh(x) = \frac{1}{2}\ln\frac{1+x}{1-x} \qquad (x\in(-1,1)).
        \]
The general solution is given by
\[ 
	y(x) = c_1 (1-x^2)^{-\frac{1}{2}} e^{\xi_1\arctanh(x)}
	+ c_2 (1-x^2)^{-\frac{1}{2}} e^{\xi_2\arctanh(x)}
\] 
where $\xi_1$ and $\xi_2$ are the two roots of the quadratic function
\[ \xi^2 +4\lambda-1=0.\]
Incidentally, it is easy to see that the finite-sample counterpart covariance matrix
is {\it doubly symmetric}, {\it i.e.},
symmetric about both the main diagonal and the secondary diagonal;
see
\cite{makhoul1981eigenvectors},
\cite{cantoni1976properties}, and
\cite{cantoni1976eigenvalues}
for more details.

If $\lambda \leq 1/4$, then
the general solution $y(x)$ is given as
\[y(x) = (1-x^2)^{-\frac{1}{2}} \left[
c_1 e^{-\theta\atanh(x)} +c_2 e^{\theta\atanh(x)}
\right]\]
where $\theta = \sqrt{1-4\lambda}$.
Using the following basic calculus results
\[
\int (1-x^2)^{-\frac{3}{2}}e^{\theta \atanh(x)} dx 
= -\frac{(\theta-x)(1-x^2)^{-\frac{1}{2}}e^{\theta \atanh(x)}}{1-\theta^2} +C,
\]
\[
\int x(1-x^2)^{-\frac{3}{2}}e^{\theta \atanh(x)} dx 
= -\frac{(\theta x-1)(1-x^2)^{-\frac{1}{2}}e^{\theta \atanh(x)}}{1-\theta^2} +C,
\]
\[
\int (1-x^2)^{-\frac{3}{2}}e^{-\theta \atanh(x)} dx 
= \frac{(\theta+x)(1-x^2)^{-\frac{1}{2}}e^{-\theta \atanh(x)}}{1-\theta^2} +C,
\]
and
\[
\int x (1-x^2)^{-\frac{3}{2}}e^{-\theta \atanh(x)} dx 
= \frac{(\theta x+1)(1-x^2)^{-\frac{1}{2}}e^{-\theta \atanh(x)}}{1-\theta^2} +C,
\]
where $C$ is a constant, we can see that
non-trivial solutions require that
\[ 
2\varepsilon (1+\theta) e^{\theta \atanh(1-2\varepsilon)}
+2\varepsilon(1-\theta) e^{-\theta \atanh(1-2\varepsilon)}=0.
	\]
Since $0\leq \theta \leq 1$,
there are no non-trivial solution if $\lambda \leq 1/4$.

If $\lambda>1/4$, then
routine algebraic operations on complex numbers lead to
the general solution $y(x)$ given as
\begin{equation}\label{eq:RnSq-sol-02}
	y(x)= \frac{1}{(1-x^2)^{\frac{1}{2}}}\left[ c_1
	\cos\left(\theta\atanh(x)\right)
	+c_2 \sin\left(\theta\atanh(x)\right)
	\right],
\end{equation}
where $\theta = 2\omega = \sqrt{4\lambda-1}$.
To find solutions satisfying the integral equation
\eqref{eq:eigensystem-01}, we can make use of the following
indefinite integrals
\[\int (1 - x^2)^{-3/2} \sin(\theta\atanh(x))dx
=\frac{-\theta \cos(\theta \atanh(x)) + x \sin(\theta \atanh(x))}
{(1 + \theta^2)\sqrt{1 - x^2}}+C,\]
\[\int (1 - x^2)^{-3/2} \cos(\theta\atanh(x))dx
=\frac{x \cos(\theta\atanh(x)) + \theta \sin(\theta \atanh(x))}
{(1 + \theta^2) \sqrt{1 - x^2}} + C,\]
\[\int x (1 - x^2)^{-3/2} \sin(\theta\atanh(x))dx
=\frac{-\theta x \cos(\theta \atanh(x)) + \sin(\theta \atanh(x))}
{(1 + \theta^2) \sqrt{1 - x^2}}+C,\]
and
\[\int x (1 - x^2)^{-3/2} \cos(\theta\atanh(x))dx
= \frac{\cos(\theta \atanh(x)) + \theta x \sin(\theta \atanh(x))}
{(1 + \theta^2) \sqrt{1 - x^2}} +C,\]
where $C$ is a constant.
Equating $f(t)$ and 
$ \lambda\int_0^1(\min(t,s)-ts)\sqrt{\psi(t)}\sqrt{\psi(s)} f(s)ds$,
with standard calculus operations, we can obtain
\[ c_1A(x) + c_2B(x) = 0,\]
where, omitting tedious details of the derivation, 
\begin{eqnarray*}
	A(x) 
	&=&
\frac{-\varepsilon \cos(\theta\atanh(1-2\varepsilon)) + \varepsilon\theta \sin(\theta \atanh(1-2\varepsilon))}
	{ \sqrt{\varepsilon(1-\varepsilon)}}
	\frac{1}{1-x^2}
\end{eqnarray*}
and 
\begin{eqnarray*}
	B(x) 
&=& 
	-\frac{\varepsilon\theta\cos(\theta\atanh(1-2\varepsilon)) + \varepsilon\sin(\theta \atanh(1-2\varepsilon))}
	{\sqrt{\varepsilon(1-\varepsilon)}}
	\frac{x}{1-x^2}.
\end{eqnarray*}
The nontrivial solutions require that with $c_1^2+c_2^2\neq 0$,
$c_1A(x) + c_2B(x)=0$ for all $x\in[-1+2\varepsilon, 1-2\varepsilon]$.
This amounts to requiring
\begin{equation}\label{eq:boundary-cond-01}
 \cos(\theta\atanh(1-2\varepsilon)) - \theta \sin(\theta \atanh(1-2\varepsilon)) = 0
\end{equation}
for $A(x)=0$, that is,
\begin{equation}\label{eq:Ax-cond}
	\tan\left(\omega \ln\frac{1-\varepsilon}{\varepsilon}\right)
	=\frac{1}{2\omega},
\end{equation}
and
\begin{equation}\label{eq:boundary-cond-02}
	\theta\cos(\theta\atanh(1-2\varepsilon)) + \sin(\theta \atanh(1-2\varepsilon)) = 0
\end{equation}
for $B(x)=0$, that is,
\begin{equation}\label{eq:Bx-cond}
	\tan\left(\omega \ln\frac{1-\varepsilon}{\varepsilon}\right)
	=-2\omega.
\end{equation}
It is easy to see that the claimed results follow 
\eqref{eq:Ax-cond} and \eqref{eq:Bx-cond}.

Regarding the claim on the presentation of the eigenvalue problem as
a Sturm-Liouville problem, 
we prove it by establishing Robin boundary conditions
for the fundamental initial conditions 
\eqref{eq:boundary-cond-01}
and
\eqref{eq:boundary-cond-02}. For notational convenience, here we 
take $a=-1+2\varepsilon$ and $b=1-2\varepsilon$.
Differentiate \eqref{eq:RnSq-sol-02} to obtain
\begin{eqnarray*}
	y'(x) & = & \;\;\; c_1 (1-x^2)^{-\frac{3}{2}}
	\left[x\cos\left( \theta\atanh(x)\right)
	-\theta\sin\left( \theta\atanh(x)\right)\right]\\
	&&
	+ c_2
	(1-x^2)^{-\frac{3}{2}}\left[x\sin\left( \theta\atanh(x)\right)
	+\theta\cos\left( \theta\atanh(x)\right)\right],
\end{eqnarray*}
which implies that
\begin{eqnarray*} 
	\left[4\varepsilon(1-\varepsilon)\right] ^{\frac{3}{2}}
	y'(a)
	&=&\;\;\; c_1\left[(-1+2\varepsilon)\cos\left( \theta\atanh(-1+2\varepsilon)\right)
	-\theta\sin\left( \theta\atanh(-1+2\varepsilon)\right)
	\right]\\
	&&
	+ c_2\left[(-1+2\varepsilon)\sin\left( \theta\atanh(-1+2\varepsilon)\right)
	+\theta\cos\left( \theta\atanh(-1+2\varepsilon)\right)
	\right]
\end{eqnarray*}
and
\begin{eqnarray*} 
	\left[4\varepsilon(1-\varepsilon)\right] ^{\frac{3}{2}}
	y'(b)
	&=&\;\;\; c_1\left[(1-2\varepsilon)\cos\left( \theta\atanh(1-2\varepsilon)\right)
	-\theta\sin\left( \theta\atanh(1-2\varepsilon)\right)
	\right]\\
	&&
	+ c_2\left[(1-2\varepsilon)\sin\left( \theta\atanh(1-2\varepsilon)\right)
	+\theta\cos\left( \theta\atanh(1-2\varepsilon)\right)
	\right].
\end{eqnarray*}
In addition, the values of $y(x)$ at the two end points are obtained
from \eqref{eq:RnSq-sol-02} as
\begin{eqnarray*} 
	\left[4\varepsilon(1-\varepsilon)\right]^{\frac{1}{2}} y(a)
&=& c_1\cos\left( \theta\atanh(-1+2\varepsilon)\right)
     + c_2\sin\left( \theta\atanh(-1+2\varepsilon)\right)
\end{eqnarray*}
and
\begin{eqnarray*}
	\left[4\varepsilon(1-\varepsilon)\right]^{\frac{1}{2}} y(b) &=& c_1\cos\left( \theta\atanh(1-2\varepsilon)\right) + c_2\sin\left( \theta\atanh(1-2\varepsilon)\right).
\end{eqnarray*}
Consider the following equivalent Robin boundary conditions
\begin{eqnarray*}
\alpha_1\left[4\varepsilon(1-\varepsilon)\right]^{\frac{1}{2}} y(a)
+ \left[4\varepsilon(1-\varepsilon)\right] ^{\frac{3}{2}}
	y'(a) &= & 0 \\
\alpha_2\left[4\varepsilon(1-\varepsilon)\right]^{\frac{1}{2}} y(b)
+ \left[4\varepsilon(1-\varepsilon)\right] ^{\frac{3}{2}}
	y'(b) &=& 0,
\end{eqnarray*}
	that is, 
\begin{eqnarray*}
0	&=& \;\;\; c_1\left[
	\;\;
	(\alpha_1 - 1 + 2\varepsilon)
	\cos\left( \theta\atanh(1-2\varepsilon)\right)
        +\theta\sin\left( \theta\atanh(1-2\varepsilon)\right)\right]\\
	&& + c_2\left[-(\alpha_1 - 1 + 2\varepsilon)\sin\left( \theta\atanh(1-2\varepsilon)\right)
        +\theta\cos\left( \theta\atanh(1-2\varepsilon)\right)\right]\\
0	&=& \;\;\; c_1\left[(\alpha_2 + 1 - 2\varepsilon)\cos\left( \theta\atanh(1-2\varepsilon)\right)
        -\theta\sin\left( \theta\atanh(1-2\varepsilon)\right)\right]\\
	&& + c_2\left[(\alpha_2 + 1 - 2\varepsilon)\sin\left( \theta\atanh(1-2\varepsilon)\right)
        +\theta\cos\left( \theta\atanh(1-2\varepsilon)\right)\right].
\end{eqnarray*}
\color{black}
For non-trivial solutions of $y(x)$, the determinant
of the matrix of the coefficients in the above system of two linear equations
must be zero.
With routine algebraic operations, it is easy to see
that this is satisfied if and only if $\alpha_1 = -2\varepsilon$ and $\alpha_2=2\varepsilon$.
This leads to the Robin boundary conditions:
\[ y(a) - 2(1-\varepsilon)y'(a)=0
\mbox{ and } y(b) + 2(1-\varepsilon)y'(b)=0.
	\]

From the Sturm-Liouville theory on the orthogonality of
the solutions to \eqref{eq:Sturm-Liouville-02}
that satisfy Robin boundary conditions, it is known that
\begin{equation}\label{eq:orth-01}
	\int_{-1+2\varepsilon}^{1-2\varepsilon}
	\frac{1}{(1-x^2)^2} g_1(x)g_2(x)dx = 0
\end{equation}
holds for any two different solutions $g_1(x)$ and $g_2(x)$.
Note that $g(x) = f(t)\psi^{-\frac{1}{2}}(t)$ with $t = (1+x)/2$.
Equation \eqref{eq:orth-01} can be written as
\[ 
	\frac{1}{2}
	\int_{-1+2\varepsilon}^{1-2\varepsilon}
	\frac{1}{4[t(1-t)]^2} \psi^{-1}(t)dt\\
	= \frac{1}{8} \int_{-1+2\varepsilon}^{1-2\varepsilon}
	f_1(t)f_2(t) dt
	= 0
\] 
This completes the proof.

\ifthenelse{1=1}{}{
\subsection{Proof of Corollary \ref{corrallary:boundary-cond}}

\color{red}
Is there a set of boundary conditions that can lead to solutions? TEST
We show that the answer is YES and the conditions are the (Robin) boundary conditions
below with $\beta_2 = 2(1-\varepsilon)$
and $\beta_1 = -2(1-\varepsilon)$.

\color{blue}
For the present eigenfunctions as the continuum limit of
eigenvectors of the finite-sample case,
we impose the following two sets of (Robin) boundary conditions:
\begin{equation}\label{eq:SL-conditions-symmetric}
        y(a) + \beta_1 y'(a)=0
        \quad\mbox{ and }\quad
        y(b) - \beta_1 y'(b)=0
\end{equation}
for the symmetric eigenvectors and
\begin{equation}\label{eq:SL-conditions-skew-symmetric}
        y(a) - \beta_2 y'(a)=0
        \quad\mbox{ and }\quad
        y(b) - \beta_2 y'(b)=0
\end{equation}
for the skew symmetric eigenvectors.\footnote{\color{red}
$\beta_1$ and $\beta_2$ must satisfy
that all the eigenfunctions are orthogonal with respective
the weight function $\psi(t)$.
}
This completes the specification of a Sturm-Liouville problem
for investigation of our asymptotic distribution of $R_n^2$.

If $\lambda<1/4$,
the general solution
$y(x)$ can be rewritten as
\[y(x) = (1-x^2)^{-\frac{1}{2}} \left[
c_1 e^{-\theta\atanh(x)} +c_2 e^{\theta\atanh(x)}
\right]\]
where $\theta = \sqrt{1-4\lambda}$.
Note that
\[y'(x) = (1-x^2)^{-\frac{3}{2}}
\left[ c_1(x-\theta) e^{-\theta\atanh(x)} +c_2(x+\theta) e^{\theta\atanh(x)} \right]
\]
For $y(x)$ to be an even function, $c_1$ and $c_2$ must satisfy $c_1=c_2$.
In this case,
\[y'(x) = c_1(1-x^2)^{-\frac{3}{2}}
\left[ (x-\theta) e^{-\theta\atanh(x)} +(x+\theta) e^{\theta\atanh(x)} \right]
\]
The boundary conditions require that at $x = a$,
\(y(a) + \beta_1y'(a)=0\), {\it i.e.},
\[
c_1\left[ (1-a^2) + \beta_1(a-\theta)\right] e^{-\theta\atanh(a)}
+c_1\left[ (1-a^2) + \beta_1(a+\theta)\right] e^{\theta\atanh(a)} =0
\]
In the $\beta_1=0$ case, this is reduced to
\[
c_1(1-a^2) \left[ e^{-\theta\atanh(a)} + e^{\theta\atanh(a)}\right] =0,
\]
leads to the only trivial solution given by $c_1=c_2=0$.\footnote{\color{red}
the general $\beta_1$ case?}

Similarly, for $y(x)$ to be an odd function, $c_1$ and $c_2$ must satisfy $c_1=-c_2$.
In this case,
\[y'(x) = c_1(1-x^2)^{-\frac{3}{2}}
\left[ (x-\theta) e^{-\theta\atanh(x)} -(x+\theta) e^{\theta\atanh(x)} \right]
\]
The boundary conditions require that at $x = a$,
\(y(a) - \beta_2y'(a)=0\), {\it i.e.},
\[
c_1\left[ (1-a^2) - \beta_2(a-\theta)\right] e^{-\theta\atanh(a)}
+c_1\left[ (1-a^2) + \beta_2(a+\theta)\right] e^{\theta\atanh(a)} =0
\]
In the $\beta_2=0$ case, this is reduced to
\[
c_1(1-a^2) \left[ e^{-\theta\atanh(a)} + e^{\theta\atanh(a)}\right] =0,
\]
leads to the only trivial solution given by $c_1=c_2=0$.\footnote{\color{red}
the general $\beta_2$ case?}


If $\lambda=1/4$, then
\[
        y(x) = (c_1+c_2)(1-x^2)^{-\frac{1}{2}}.
\]
which is symmetric with
\[
        y'(x) 
         = (c_1+c_2)x(1-x^2)^{-\frac{3}{2}}.
\]
\[
        y(a) + \beta_1y'(a) = (c_1+c_2)(1-a^2)^{-\frac{3}{2}}
        \left[ (1-a^2) +\beta_1 a
         \right] =0.
\]
This implies that
$c_1+c_2\neq 0$
and that $\beta_1$ must be given as
\[ \beta_1 = -\frac{4\varepsilon(1-\varepsilon)}{2\varepsilon-1}
 = \frac{4\varepsilon(1-\varepsilon)}{1-2\varepsilon}.\]
When this $\beta_1$ is taken,
this case provides as a limiting case of one eigenfunction
in the $\lambda>1/4$ case.

If $\lambda>1/4$, then
routine algebraic operations on complex numbers leads to
the general solution $y(x)$ given as
\begin{equation}\label{eq:RnSq-sol-02}
	y(x)= \frac{1}{(1-x^2)^{\frac{1}{2}}}\left[ c_1
	\cos\left(\omega\ln\frac{1+x}{1-x}\right)
	+c_2 \sin\left(\omega\ln\frac{1+x}{1-x}\right)
	\right],
\end{equation}
where $\omega = \frac{1}{2}\sqrt{4\lambda-1}$.
Thus, the boundary conditions for the symmetric eigenfunctions
require $c_2=0$ and $c_1\neq 0$, which leads to
\begin{equation}\label{eq:SL-conditions-symmetric-solution}
	y_1(x)= c_1(1-x^2)^{-\frac{1}{2}}
	\cos\left(\omega\ln\frac{1+x}{1-x}\right).
\end{equation}
If $\beta_1=0$, then
\[
	\cos\left(\omega\ln\frac{1-\varepsilon}{\varepsilon}\right) = 0.
	\]
In this case,
\[ \omega_k = \frac{k\pi+\frac{\pi}{2}}{\ln\frac{1-\varepsilon}{\varepsilon}}
\qquad(k=0, 1, 2,...).
	\]
Otherwise,
from the first-order derivate of $y_1(x)$,
\begin{eqnarray*}
	y_1'(x) &=&
	c_1x (1-x^2)^{-\frac{3}{2}}\cos\left(\omega\ln\frac{1+x}{1-x}\right)
	-c_1 2\omega  (1-x^2)^{-\frac{3}{2}}\sin\left(\omega\ln\frac{1+x}{1-x}\right),
\end{eqnarray*}
we have the boundary condition, provided that $c_1\neq 0$
for non-trivial solutions, $w_k$, given by
\begin{equation}\label{eq:SL-conditions-symmetric-solution-02}
	\tan\left(\omega_k\ln\frac{1-\varepsilon}{\varepsilon}\right)
	= \frac{\left[\frac{1-2\varepsilon}{2} - \frac{2\varepsilon(1-\varepsilon)
	}{\beta_1}
	\right] \ln\frac{1-\varepsilon}{\varepsilon}
	}{\omega_k\ln\frac{1-\varepsilon}{\varepsilon}} 
\end{equation}
with
\[ k\pi \leq \omega_k < k\pi+\frac{\pi}{2}
	\]
	for $k=0, 1, 2, ...$

Similarly, the boundary conditions for the skew symmetric eigenfunctions
require that $c_1=0$ and $c_2\neq 0$, which leads to
\begin{equation}\label{eq:SL-conditions-symmetric-solution}
	y_2(x)= c_2(1-x^2)^{-\frac{1}{2}}
	\sin\left(\omega\ln\frac{1+x}{1-x}\right).
\end{equation}
If $\beta_2=0$, then
\[
	\sin\left(\omega\ln\frac{1-\varepsilon}{\varepsilon}\right) = 0.
	\]
In this case,
\[ \omega_k = \frac{k\pi}{\ln\frac{1-\varepsilon}{\varepsilon}}
\qquad(k=1, 2,...).
	\]
Otherwise,
from the first-order derivate of $y_2(x)$,
\begin{eqnarray*}
	y_2'(x) &=&
	c_2x (1-x^2)^{-\frac{3}{2}}\sin\left(\omega\ln\frac{1+x}{1-x}\right)
	+c_2 2\omega (1-x^2)^{-\frac{3}{2}}\cos\left(\omega\ln\frac{1+x}{1-x}\right),
\end{eqnarray*}
we have the boundary condition, provided that $c_2\neq 0$
for non-trivial solutions,
$w_k$, given by
\begin{equation}\label{eq:SL-conditions-skew-symmetric-solution-03}
 \cot\left(\omega\ln\frac{1-\varepsilon}{\varepsilon}\right)
	= \frac{\left[  \frac{- 4\varepsilon(1-\varepsilon) + \beta_2(2\varepsilon-1)}{2\beta_2}
	\right]\ln\frac{1-\varepsilon}{\varepsilon}
}{\omega\ln\frac{1-\varepsilon}{\varepsilon}}
\end{equation}
with
\[(k-1)\pi+\frac{\pi}{2}\leq \omega_k < k\pi
\qquad(k=1,2,...).\]
\footnote{\color{red}
What would be the solutions that provide
better approximations to the numerical results?
But it is pointed out that in ???
The numerical results are not accurate for finding
eigenvalues.
}

\footnote{\color{red}
A possible approximation is to find $\omega$'s:
	$ y(A - \frac{1}{\lambda} I) y = 0$
...
}

\color{black}

}



\subsection{Proof of Theorem \ref{thm:AD-limit-kernel}}
\label{proof:Wn}

Due to the symmetry property of $\kappa(t,s)$, it is sufficient to consider the case $s\geq t$.
In this case, the element in the continuum limit with
$i/(n+1)\rightarrow t$ and $j/(n+1)\rightarrow s$, where $0<t < s < 1$, is obtained as follows.
	Note that $n+1-(j-i)>0$,
\begin{eqnarray}
	&&\lim_{n\rightarrow \infty} \frac{1}{(n+1)^2} 
	\left[ \sum_{k=j-i}^n\psi_k[k-(j-i)]
	+\sum_{k=n+1-(j-i)}^n\psi_k[k-i+j-(n+1)]
	-\sum_{k=1}^n\psi_k \frac{k^2}{n+1}\right]
	\nonumber\\
&=& \lim_{n\rightarrow \infty} \left[\sum_{k=j-i}^n\frac{ k-(j-i)}{k(n+1-k)} +\sum_{k=n+1-(j-i)}^n 	\frac{ k-i+j-(n+1)}{k(n+1-k)} -\frac{1}{n+1}\sum_{k=1}^n \frac{k}{n+1-k} \right]\nonumber \\
\ifthenelse{\boolean{shortversion}}{}{
&=& \color{red} \lim_{n\rightarrow \infty}
	\left[ \sum_{k=j-i}^n\frac{1}{n+1-k} 
		-\frac{j-i}{n+1}
		\sum_{k=j-i}^n\left( \frac{1}{k} +\frac{1}{n+1-k} \right)
	+\sum_{k=n+1-(j-i)}^n 	\frac{ 1}{n+1-k}
	\right.\\
	&&\left.\;\;\;\;\;
	-\left(1-\frac{j-i}{n+1}\right)
	\sum_{k=n+1-(j-i)}^n \left( \frac{1}{k} +\frac{1}{n+1-k} \right)
	-\frac{1}{n+1}\sum_{k=1}^n \frac{n+1-k}{k}
	\right]
	\\
&=& \color{magenta} \lim_{n\rightarrow \infty}
	\left[ \left(1-\frac{j-i}{n+1}\right)
		\sum_{k=j-i}^n\frac{1}{n+1-k} 
		-\frac{j-i}{n+1} \sum_{k=j-i}^n \frac{1}{k}
	+\frac{j-i}{n+1}\sum_{k=n+1-(j-i)}^n 	\frac{ 1}{n+1-k}
	\right.\\
	&&\left.\;\;\;\;\;
	-\left(1-\frac{j-i}{n+1}\right)
	\sum_{k=n+1-(j-i)}^n \frac{1}{k}
	-\sum_{k=1}^n \frac{1}{k} +\frac{n}{n+1} \right]
	\\
&=& \color{blue} \lim_{n\rightarrow \infty} 
	\left[ \left(1-\frac{j-i}{n+1}\right)
		\sum_{k=1}^{n+1-(j-i)}\frac{1}{k} 
		-\frac{j-i}{n+1} \sum_{k=j-i}^n \frac{1}{k}
	+\frac{j-i}{n+1}\sum_{k=1}^{j-i} 	\frac{ 1}{k}
	\right.\\
	&&\left.\;\;\;\;\;
	-\left(1-\frac{j-i}{n+1}\right)
	\sum_{k=n+1-(j-i)}^n \frac{1}{k}
	-\sum_{k=1}^n \frac{1}{k} +\frac{n}{n+1} \right]
	\\
&=& \color{red} \lim_{n\rightarrow \infty}
	\left[ \sum_{k=1}^{n+1-(j-i)}\frac{1}{k} 
		-\frac{j-i}{n+1} \sum_{k=1}^{n+1-(j-i)}\frac{1}{k} 
		-\frac{j-i}{n+1} \sum_{k=j-i}^n \frac{1}{k}
	+\frac{j-i}{n+1}\sum_{k=1}^{j-i} 	\frac{ 1}{k}
	\right.\\
	&&\left.\;\;\;\;\;
	-\sum_{k=n+1-(j-i)}^n \frac{1}{k}
	+\frac{j-i}{n+1}\sum_{k=n+1-(j-i)}^n \frac{1}{k}
	-\sum_{k=1}^n \frac{1}{k} +\frac{n}{n+1} \right]
	\\
}
&=& \lim_{n\rightarrow \infty}
	\left[
		-\frac{j-i}{n+1} \sum_{k=1}^{n+1-(j-i)}\frac{1}{k} 
		-\frac{j-i}{n+1} \sum_{k=j-i}^n \frac{1}{k}
	+\frac{j-i}{n+1}\sum_{k=1}^{j-i} 	\frac{ 1}{k}
	+\frac{j-i}{n+1}\sum_{k=n+1-(j-i)}^n \frac{1}{k}
	\right.\nonumber\\
	&&\left.\hspace{0.5in}
+		\sum_{k=1}^{n+1-(j-i)}\frac{1}{k} 
	-\sum_{k=n+1-(j-i)}^n \frac{1}{k}
	-\sum_{k=1}^n \frac{1}{k} +\frac{n}{n+1} \right].
\label{eq:AD-proof-01}
\end{eqnarray}
If $j-i\ge (n+1)/2$,
that is, $j-i \ge n+1 - (j-i)$ and $s-t \ge \frac{1}{2}$,
then \eqref{eq:AD-proof-01} becomes
\begin{eqnarray*}
\ifthenelse{\boolean{shortversion}}{
	&&
}{
&=& \color{cyan} \lim_{n\rightarrow \infty}
	\left[
	\frac{j-i}{n+1}\sum_{k=n+1-(j-i)+1}^{j-i} 	\frac{ 1}{k}
	+\frac{j-i}{n+1}\sum_{k=n+1-(j-i)}^{j-i} \frac{1}{k}
	-\sum_{k=n+1-(j-i)}^n \frac{1}{k}
	-\sum_{k=n+1-(j-i)+1}^n \frac{1}{k} +\frac{n}{n+1} \right]
	\\
&=&
}
	\lim_{n\rightarrow \infty}
	\left[
	\frac{2(j-i)}{n+1}\sum_{k=n+1-(j-i)+1}^{j-i} 	\frac{ 1}{k}
	+\left(\frac{j-i}{n+1}
	-1\right)\frac{1}{n+1-(j-i)}
	-2\sum_{k=n+1-(j-i)+1}^n \frac{1}{k} +\frac{n}{n+1} \right]
	\\
&=& 2(s-t)\lim_{n\rightarrow \infty} \left[\ln(j-i) - \ln(n+1-(j-i))\right]
	-2\lim_{n\rightarrow \infty}\left[\ln(n)-\ln(n+1-(j-i))\right] +1
	\\
\ifthenelse{\boolean{shortversion}}{}{
&=& 2(s-t)\lim_{n\rightarrow \infty}
	\ln\frac{j-i}{n+1-(j-i)}
	-2\lim_{n\rightarrow \infty} \ln\frac{n}{n+1-(j-i)} +1
	\\
}
&=& 2(s-t) \ln\frac{s-t}{1-(s-t)} -2\ln\frac{1}{1-(s-t)} +1
\\
\ifthenelse{\boolean{shortversion}}{}{
&=& \color{blue} 2(s-t) \ln(s-t) - 2(s-t) \ln(1-(s-t)) +2\ln(1-(s-t))+1
\\
}
	&=& 2\left[(s-t) \ln(s-t) +[1 - (s-t)] \ln(1-(s-t))\right]+1,
\end{eqnarray*}
where the method of series estimation with integrals is used.
If $j-i < (n+1)/2$,
that is, $j-i < n+1 - (j-i)$ and $s-t < \frac{1}{2}$,
then \eqref{eq:AD-proof-01} becomes
\begin{eqnarray*}
\ifthenelse{\boolean{shortversion}}{ &&}{
&=& \color{cyan} \lim_{n\rightarrow \infty}
	\left[ 
		-\frac{j-i}{n+1} \sum_{k=1}^{n+1-(j-i)}\frac{1}{k} 
		-\frac{j-i}{n+1} \sum_{k=j-i}^n \frac{1}{k}
	+\frac{j-i}{n+1}\sum_{k=1}^{j-i} 	\frac{ 1}{k}
	+\frac{j-i}{n+1}\sum_{k=n+1-(j-i)}^n \frac{1}{k}
	\right.\\
	&&\left.\;\;\;\;\;
+		\sum_{k=1}^{n+1-(j-i)}\frac{1}{k} 
	-\sum_{k=n+1-(j-i)}^n \frac{1}{k}
	-\sum_{k=1}^n \frac{1}{k} +\frac{n}{n+1} \right]
	\\
&=&
}
	\lim_{n\rightarrow \infty}
	\left[ -\frac{j-i}{n+1} \sum_{k=j-i+1}^{n+1-(j-i)}\frac{1}{k} 
		-\frac{j-i}{n+1} \sum_{k=j-i}^{n+1-(j-i)-1} \frac{1}{k}
	-\sum_{k=n+1-(j-i)}^n \frac{1}{k}
	-\sum_{k=n+1-(j-i)+1}^n \frac{1}{k} +\frac{n}{n+1} \right]
	\\
&=& -2(s-t)\lim_{n\rightarrow \infty}\ln\frac{n+1-(j-i)}{j-i}
	 -2\lim_{n\rightarrow \infty}\ln\frac{n}{n+1-(j-i)}
	+1 \\
&=& -2(s-t)\ln\frac{1-(s-t)}{s-t}
	 -2\ln\frac{1}{1-(s-t)} +1 \\
\ifthenelse{\boolean{shortversion}}{ }{
&=& \color{blue}
	-2(s-t)\ln(1-(s-t)) +2(s-t)\ln(s-t)
	 +2\ln(1-(s-t)) +1 \\
}
&=& 2\left[(1-(s-t))\ln(1-(s-t)) +(s-t)\ln(s-t)\right] +1,
\end{eqnarray*}
where the method of series estimation with integrals is used again.
For the $s=t=i/(n+1)$ case, it is easy to see that
a simplified version of the above derivation 
gives the claimed result.
In summary, we have Equation \eqref{eq:Wn-asym-kernel} and
complete the proof.

\subsection{Proof of Theorem \ref{thm:New-limit-kernel}}
\label{proof:Rn}

Due to the symmetry property of $\kappa(t,s)$, it is sufficient to consider the case $s\geq t$.
In this, the element in the continuum limit with
$t=i/(n+1)$ and $s=j/(n+1)$, where $0<t < s < 1$, is obtained as follows:
	\color{black}
\begin{eqnarray}
	&&\lim_{n\rightarrow \infty}
	\frac{1}{(n+1)^2}
	\left[\sum_{k=j-i}^n\psi_k[k-(j-i)]
	+\sum_{k=n+1-(j-i)}^n\psi_k[k-i+j-(n+1)]
	-\sum_{k=1}^n\psi_k\frac{k^2}{n+1}
	\right]
	\nonumber\\
&=& \lim_{n\rightarrow \infty} (n+1)^2
	\left[
	\sum_{k=j-i}^n\frac{k-(j-i)}{[k(n+1-k)]^2}
	+\sum_{k=n+1-(j-i)}^n\frac{k-i+j-(n+1)}{[k(n+1-k)]^2}
	-\frac{1}{n+1}\sum_{k=1}^n\frac{1}{(n+1-k)^2}
	\right]\nonumber\\
\ifthenelse{\boolean{shortversion}}{ }{
	&=& \color{blue} \lim_{n\rightarrow \infty} (n+1)
	\left[ \sum_{k=j-i}^n\frac{1}{n+1-k}\left[
			\frac{1}{k}+\frac{1}{n+1-k}\right]
		-\frac{j-i}{n+1}\sum_{k=j-i}^n
		\left[\frac{1}{k}+\frac{1}{n+1-k}\right]^2
	\right. \nonumber  \\
	&& \left.
	+\sum_{k=n+1-(j-i)}^n\frac{1}{n+1-k}
	\left[ \frac{1}{k} + \frac{1}{n+1-k} \right]
	-\left(1-\frac{j-i}{n+1}\right)
	\sum_{k=n+1-(j-i)}^n\left[\frac{1}{k}+\frac{1}{n+1-k}\right]^2
	\color{red} -\sum_{k=1}^n\frac{1}{k^2}
	\right]\nonumber\\
}
	&=& \lim_{n\rightarrow \infty} (n+1)
	\left[ \sum_{\ell \in \{j-i, n+1-(j-i)\}} \sum_{k=\ell}^n\frac{1}{n+1-k}\left[
			\frac{1}{k}+\frac{1}{n+1-k}\right]
		-\frac{j-i}{n+1}\sum_{k=j-i}^n
		\left[\frac{1}{k}+\frac{1}{n+1-k}\right]^2
	\right.  \nonumber\\
	&& \left.
	- \sum_{k=n+1-(j-i)}^n\left[\frac{1}{k}+\frac{1}{n+1-k}\right]^2
	+\frac{j-i}{n+1}
	\sum_{k=n+1-(j-i)}^n\left[\frac{1}{k}+\frac{1}{n+1-k}\right]^2
	 -\sum_{k=1}^n\frac{1}{k^2}
	\right].
\label{eq:Optim-proof-01}
\end{eqnarray}
If $j-i < (n+1)/2$, that is,
	$j-i < n+1 - (j-i)$ and $s-t < \frac{1}{2}$,
then \eqref{eq:Optim-proof-01} becomes
\begin{eqnarray*}
\ifthenelse{\boolean{shortversion}}{ && }{
&=& \color{red} \lim_{n\rightarrow \infty} (n+1)
	\left[ \sum_{\ell \in \{j-i, n+1-(j-i)\}} \sum_{k=\ell}^n\frac{1}{n+1-k}\left[
			\frac{1}{k}+\frac{1}{n+1-k}\right]
		-\frac{j-i}{n+1}\sum_{k=j-i}^{n+1-(j-1)-1}
		\left[\frac{1}{k}+\frac{1}{n+1-k}\right]^2
	\right.  \\
	&& \left.
	- \sum_{k=n+1-(j-i)}^n\frac{1}{k}\left[\frac{1}{k}+\frac{1}{n+1-k}\right]
	- \sum_{k=n+1-(j-i)}^n\frac{1}{n+1-k}\left[\frac{1}{k}+\frac{1}{n+1-k}\right]
	\color{red} -\sum_{k=1}^n\frac{1}{k^2}
	\right]\\
&=&}
	\lim_{n\rightarrow \infty} (n+1)
	\left[
		\sum_{k=j-i}^{n+1-(j-i)-1}\frac{1}{n+1-k}\frac{1}{k}
		-\frac{j-i}{n+1}\sum_{k=j-i}^{n+1-(j-1)-1}
		\left[\frac{1}{k}+\frac{1}{n+1-k}\right]^2
	-2 \sum_{n+1-(j-i)}^n\frac{1}{k^2}
	\right]\\
&=&  2 \ln\frac{1-(s-t)}{s-t}
	-(s-t)\int_{s-t}^{1-(s-t)}
	\left(\frac{1}{x}+\frac{1}{1-x}\right)^2 dx
	- 2\int_{1-(s-t)}^1\frac{1}{x^2}dx
	\\
\ifthenelse{\boolean{shortversion}}{ }{
&=& 2 \ln\frac{1-(s-t)}{s-t}
	-4(s-t) \left[ \frac{\frac{1}{2}-(s-t)}{(s-t)[1-(s-t)]}
		+ \ln\frac{1-(s-t)}{s-t}
		\right] - 2\frac{s-t}{1-(s-t)}.
		\\
&=& -2 \ln\frac{s-t}{1-(s-t)} - 2\frac{s-t}{1-(s-t)}
	+4(s-t) \left[ \frac{(s-t)-\frac{1}{2}}{(s-t)[1-(s-t)]}
		+ \ln\frac{s-t}{1-(s-t)}
		\right]
		\\
&=& -2 \ln\frac{s-t}{1-(s-t)} - 2\frac{s-t}{1-(s-t)}
	+4\frac{(s-t)-\frac{1}{2}}{1-(s-t)}
		+ 4(s-t) \ln\frac{s-t}{1-(s-t)}
		\\
&=&  \frac{-2(s-t)}{1-(s-t)} +\frac{4(s-t)-2}{1-(s-t)}
		+ 4\left[(s-t) -\frac{1}{2}\right] \ln\frac{s-t}{1-(s-t)}
		\\
}
&=&  4\left[(s-t) -\frac{1}{2}\right] \ln\frac{s-t}{1-(s-t)} -2,
\end{eqnarray*}
where the method of series estimation with integrals is used.
If $j-i\geq (n+1)/2$, that is,
	$j-i\geq n+1 - (j-i)$ and $s-t\geq \frac{1}{2}$,
then \eqref{eq:Optim-proof-01} becomes
\begin{eqnarray*}
\ifthenelse{\boolean{shortversion}}{ && }{
	&=& (n+1) \color{red} \lim_{n\rightarrow \infty} 
	\left[ \sum_{\ell \in \{j-i, n+1-(j-i)\}} \sum_{k=\ell}^n\frac{1}{n+1-k}\left[
			\frac{1}{k}+\frac{1}{n+1-k}\right]
		- \sum_{k=n+1-(j-i)}^n\frac{1}{k}\left[\frac{1}{k}+\frac{1}{n+1-k}\right]
	\right.  \\
	&& \left.
	- \sum_{k=n+1-(j-i)}^n\frac{1}{n+1-k}\left[\frac{1}{k}+\frac{1}{n+1-k}\right]
	+\frac{j-i}{n+1}
	\sum_{k=n+1-(j-i)}^{j-i-1}\left[\frac{1}{k}+\frac{1}{n+1-k}\right]^2
	\color{red} -\sum_{k=1}^n\frac{1}{k^2}
	\right]\\
	&=& (n+1) \color{blue} \lim_{n\rightarrow \infty}
	\left[ \sum_{k=j-i}^n\frac{1}{n+1-k}\left[
			\frac{1}{k}+\frac{1}{n+1-k}\right]
		- \sum_{k=1}^{j-i}\frac{1}{n+1-k}\left[\frac{1}{k}+\frac{1}{n+1-k}\right]
	\right.  \\
	&& \left.
	+\frac{j-i}{n+1}
	\sum_{k=n+1-(j-i)}^{j-i-1}\left[\frac{1}{k}+\frac{1}{n+1-k}\right]^2
	\color{red} -\sum_{k=1}^n\frac{1}{k^2}
	\right]\\
&=&
	}
	\lim_{n\rightarrow \infty} 
	(n+1)
	\left[ -\frac{2}{n+1} \sum_{k=n+1-(j-i)+1}^{j-i}\frac{1}{n+1-k}
		-2\sum_{k=1}^{j-i}\frac{1}{(n+1-k)^2}
	\right.\\ && \left. \hspace{1.0in}
	+\frac{j-i}{n+1}
	\sum_{k=n+1-(j-i)}^{j-i-1}\left(\frac{1}{k}+\frac{1}{n+1-k}\right)^2
	\right]\\
\ifthenelse{\boolean{shortversion}}{ }{
&=&  -2 \ln\frac{s-t}{1-(s-t)}
		-2\lim_{n\rightarrow \infty}\frac{1}{n+1}\sum_{k=1}^{j-i}\frac{1}{\left(1-\frac{k}{n+1}\right)^2}
	+(s-t)\lim_{n\rightarrow \infty}\frac{1}{n+1}
	\sum_{k=n+1-(j-i)}^{j-i-1}\left[\frac{n+1}{k}+\frac{n+1}{n+1-k}\right]^2
\\
}
&=& -2 \ln\frac{s-t}{1-(s-t)}
		-2\int_0^{s-t}\frac{1}{\left(1-x\right)^2}dx
	+(s-t)
	\int_{1-(s-t)}^{s-t}\left(\frac{1}{x}+\frac{1}{1-x}\right)^2dx
	\\
\ifthenelse{\boolean{shortversion}}{ }{
&=&  -2 \ln\frac{s-t}{1-(s-t)}
	-2\frac{s-t}{1-(s-t)}
	+4(s-t)
	\left[\frac{s-t-\frac{1}{2}}{(s-t)[1-(s-t)]}
		+
		\ln\frac{s-t}{1-(s-t)}
		\right]\\
}
&=&  4\left[(s-t) -\frac{1}{2}\right] \ln\frac{s-t}{1-(s-t)} -2,
\end{eqnarray*}
the same as in the $s-t < \frac{1}{2}$ case,
where the method of series estimation with integrals is used again.
For the $s=t=i/(n+1)$ case, it is easy to see that
a simplified version of the above derivation 
gives the claimed result.
In summary, we have Equation \eqref{eq:Rn-asymp-kernal} and
complete the proof.

\pagebreak

\pagebreak

\pagebreak

\pagebreak

\pagebreak

\pagebreak

\pagebreak

\pagebreak

\clearpage

\setlength{\tabcolsep}{2pt}
\vspace{1in}

\begin{table}[!htp]
\centering
\footnotesize

\caption{
        Power comparison at level 0.05 for the case in Section \ref{s:numerical} with $\{\tau, (\eta-\sigma, \eta+\sigma)\} = \{3, (0, 0.1)\}$, where $\mbox{CS}_1$ and $\mbox{CS}_2$ are the corresponding circularized versions defined in
        \eqref{eq:CS-mean-T} and \eqref{eq:CS-max-T}.
        }
        \label{tbl:pow-1}
\begin{tabular}{rcrcrcrcrcrcrcrcrcrcrcrcrcrcrcrcrcrcr}
\multicolumn{37}{c}{}\\
\hline\hline
\multicolumn{37}{c}{}\\
$n$&& \multicolumn{5}{c}{$W_n^2$}&& \multicolumn{5}{c}{$R_n^2$}
&& \multicolumn{5}{c}{AD}&& \multicolumn{5}{c}{LR}
&& \multicolumn{5}{c}{CvM}&& \multicolumn{5}{c}{KS}\\
\cline{1-1}\cline{3-7}\cline{9-13}\cline{15-19}\cline{21-25}\cline{27-31}\cline{33-37}
&& && $\mbox{CS}_{1}$ && $\mbox{CS}_{2}$&& && $\mbox{CS}_{1}$ && $\mbox{CS}_{2}$&& && $\mbox{CS}_{1}$ && $\mbox{CS}_{2}$&&&& $\mbox{CS}_{1}$ && $\mbox{CS}_{2}$&&&& $\mbox{CS}_{1}$ && $\mbox{CS}_{2}$&& && $\mbox{CS}_{1}$ && $\mbox{CS}_{2}$\\
\cline{5-5}\cline{7-7}
\cline{11-11}\cline{13-13}
\cline{17-17}\cline{19-19}
\cline{23-23}\cline{25-25}
\cline{29-29}\cline{31-31}
\cline{35-35}\cline{37-37}\multicolumn{37}{c}{}\\
10 &&  0.19 &&  0.17 &&  0.21 &&  0.21 &&  0.19 &&  0.21 &&  0.15 &&  0.16 &&  0.19 &&  0.13 &&  0.18 &&  0.17 &&  0.07 &&  0.11 &&  0.14 &&  0.08 &&  0.12 &&  0.13 \\
50 &&  0.46 &&  0.39 &&  0.76 &&  0.60 &&  0.60 &&  0.95 &&  0.39 &&  0.40 &&  0.76 &&  0.41 &&  0.61 &&  0.99 &&  0.11 &&  0.28 &&  0.63 &&  0.32 &&  0.50 &&  0.52 \\
100 &&  0.79 &&  0.78 &&  0.99 &&  0.90 &&  0.97 &&  1.00 &&  0.74 &&  0.76 &&  0.99 &&  0.78 &&  0.97 &&  1.00 &&  0.19 &&  0.55 &&  0.96 &&  0.69 &&  0.90 &&  0.92 \\
150 &&  0.95 &&  0.96 &&  1.00 &&  0.98 &&  1.00 &&  1.00 &&  0.93 &&  0.96 &&  1.00 &&  0.95 &&  1.00 &&  1.00 &&  0.28 &&  0.82 &&  1.00 &&  0.93 &&  1.00 &&  1.00 \\
200 &&  0.99 &&  1.00 &&  1.00 &&  1.00 &&  1.00 &&  1.00 &&  0.99 &&  1.00 &&  1.00 &&  0.99 &&  1.00 &&  1.00 &&  0.43 &&  0.96 &&  1.00 &&  0.99 &&  1.00 &&  1.00 \\
250 &&  1.00 &&  1.00 &&  1.00 &&  1.00 &&  1.00 &&  1.00 &&  1.00 &&  1.00 &&  1.00 &&  1.00 &&  1.00 &&  1.00 &&  0.59 &&  1.00 &&  1.00 &&  1.00 &&  1.00 &&  1.00 \\
300 &&  1.00 &&  1.00 &&  1.00 &&  1.00 &&  1.00 &&  1.00 &&  1.00 &&  1.00 &&  1.00 &&  1.00 &&  1.00 &&  1.00 &&  0.77 &&  1.00 &&  1.00 &&  1.00 &&  1.00 &&  1.00 \\
\multicolumn{37}{c}{}\\
\hline
\hline
\end{tabular}
\end{table}

\vspace{1in}

\setlength{\tabcolsep}{2pt}
\begin{table}[!htp]
\centering
\footnotesize
\caption{
        Power comparison at level 0.05 for the case in Section \ref{s:numerical} with $\{\tau, (\eta-\sigma, \eta+\sigma)\} =\{3, (0.45, 0.55)\}$,
        where $\mbox{CS}_1$ and $\mbox{CS}_2$ are the corresponding circularized versions defined in
        \eqref{eq:CS-mean-T} and \eqref{eq:CS-max-T}.
        }
        \label{tbl:pow-2}
\begin{tabular}{rcrcrcrcrcrcrcrcrcrcrcrcrcrcrcrcrcrcr}
\multicolumn{37}{c}{}\\
\hline\hline
\multicolumn{37}{c}{}\\
$n$&& \multicolumn{5}{c}{$W_n^2$}&& \multicolumn{5}{c}{$R_n^2$}
&& \multicolumn{5}{c}{AD}&& \multicolumn{5}{c}{LR}
&& \multicolumn{5}{c}{CvM}&& \multicolumn{5}{c}{KS}\\
\cline{1-1}\cline{3-7}\cline{9-13}\cline{15-19}\cline{21-25}\cline{27-31}\cline{33-37}
&& && $\mbox{CS}_{1}$ && $\mbox{CS}_{2}$&& && $\mbox{CS}_{1}$ && $\mbox{CS}_{2}$&& && $\mbox{CS}_{1}$ && $\mbox{CS}_{2}$&&&& $\mbox{CS}_{1}$ && $\mbox{CS}_{2}$&&&& $\mbox{CS}_{1}$ && $\mbox{CS}_{2}$&& && $\mbox{CS}_{1}$ && $\mbox{CS}_{2}$\\
\cline{5-5}\cline{7-7}
\cline{11-11}\cline{13-13}
\cline{17-17}\cline{19-19}
\cline{23-23}\cline{25-25}
\cline{29-29}\cline{31-31}
\cline{35-35}\cline{37-37}\multicolumn{37}{c}{}\\
10 &&  0.05 &&  0.08 &&  0.10 &&  0.06 &&  0.09 &&  0.12 &&  0.05 &&  0.08 &&  0.10 &&  0.05 &&  0.09 &&  0.11 &&  0.06 &&  0.07 &&  0.07 &&  0.08 &&  0.08 &&  0.07 \\
50 &&  0.07 &&  0.34 &&  0.71 &&  0.06 &&  0.54 &&  0.94 &&  0.08 &&  0.32 &&  0.69 &&  0.06 &&  0.53 &&  0.99 &&  0.10 &&  0.23 &&  0.54 &&  0.32 &&  0.42 &&  0.43 \\
100 &&  0.13 &&  0.73 &&  0.99 &&  0.09 &&  0.96 &&  1.00 &&  0.13 &&  0.73 &&  0.99 &&  0.08 &&  0.96 &&  1.00 &&  0.17 &&  0.52 &&  0.95 &&  0.65 &&  0.88 &&  0.91 \\
150 &&  0.19 &&  0.95 &&  1.00 &&  0.11 &&  1.00 &&  1.00 &&  0.20 &&  0.95 &&  1.00 &&  0.10 &&  1.00 &&  1.00 &&  0.27 &&  0.80 &&  1.00 &&  0.87 &&  1.00 &&  1.00 \\
200 &&  0.26 &&  1.00 &&  1.00 &&  0.13 &&  1.00 &&  1.00 &&  0.30 &&  1.00 &&  1.00 &&  0.13 &&  1.00 &&  1.00 &&  0.39 &&  0.95 &&  1.00 &&  0.97 &&  1.00 &&  1.00 \\
250 &&  0.40 &&  1.00 &&  1.00 &&  0.16 &&  1.00 &&  1.00 &&  0.37 &&  1.00 &&  1.00 &&  0.14 &&  1.00 &&  1.00 &&  0.52 &&  1.00 &&  1.00 &&  0.99 &&  1.00 &&  1.00 \\
300 &&  0.52 &&  1.00 &&  1.00 &&  0.21 &&  1.00 &&  1.00 &&  0.49 &&  1.00 &&  1.00 &&  0.16 &&  1.00 &&  1.00 &&  0.64 &&  1.00 &&  1.00 &&  1.00 &&  1.00 &&  1.00 \\
\multicolumn{37}{c}{}\\
\hline
\hline
\end{tabular}
\end{table}

\vspace{1in}

\setlength{\tabcolsep}{2pt}
\begin{table}[!htp]
\centering
\footnotesize
\caption{
        Power comparison at level 0.05 for the case in Section \ref{s:numerical} with $\{\tau, (\eta-\sigma, \eta+\sigma)\} = \{ 0.75, (0, 0.5)\}$,
        where $\mbox{CS}_1$ and $\mbox{CS}_2$ are the corresponding circularized versions defined in
        \eqref{eq:CS-mean-T} and \eqref{eq:CS-max-T}.
        }
        \label{tbl:pow-3}
\begin{tabular}{rcrcrcrcrcrcrcrcrcrcrcrcrcrcrcrcrcrcr}
\multicolumn{37}{c}{}\\
\hline\hline
\multicolumn{37}{c}{}\\
$n$&& \multicolumn{5}{c}{$W_n^2$}&& \multicolumn{5}{c}{$R_n^2$}
&& \multicolumn{5}{c}{AD}&& \multicolumn{5}{c}{LR}
&& \multicolumn{5}{c}{CvM}&& \multicolumn{5}{c}{KS}\\
\cline{1-1}\cline{3-7}\cline{9-13}\cline{15-19}\cline{21-25}\cline{27-31}\cline{33-37}
&& && $\mbox{CS}_{1}$ && $\mbox{CS}_{2}$&& && $\mbox{CS}_{1}$ && $\mbox{CS}_{2}$&& && $\mbox{CS}_{1}$ && $\mbox{CS}_{2}$&&&& $\mbox{CS}_{1}$ && $\mbox{CS}_{2}$&&&& $\mbox{CS}_{1}$ && $\mbox{CS}_{2}$&& && $\mbox{CS}_{1}$ && $\mbox{CS}_{2}$\\
\cline{5-5}\cline{7-7}
\cline{11-11}\cline{13-13}
\cline{17-17}\cline{19-19}
\cline{23-23}\cline{25-25}
\cline{29-29}\cline{31-31}
\cline{35-35}\cline{37-37}\multicolumn{37}{c}{}\\
10 &&  0.18 &&  0.23 &&  0.20 &&  0.15 &&  0.23 &&  0.20 &&  0.15 &&  0.22 &&  0.21 &&  0.09 &&  0.22 &&  0.20 &&  0.13 &&  0.22 &&  0.24 &&  0.15 &&  0.22 &&  0.24 \\
50 &&  0.52 &&  0.77 &&  0.82 &&  0.38 &&  0.81 &&  0.82 &&  0.50 &&  0.75 &&  0.82 &&  0.26 &&  0.80 &&  0.83 &&  0.39 &&  0.75 &&  0.83 &&  0.60 &&  0.80 &&  0.81 \\
100 &&  0.84 &&  0.98 &&  0.99 &&  0.68 &&  0.99 &&  0.99 &&  0.81 &&  0.98 &&  1.00 &&  0.52 &&  0.99 &&  0.99 &&  0.76 &&  0.98 &&  0.99 &&  0.92 &&  0.99 &&  0.99 \\
150 &&  0.97 &&  1.00 &&  1.00 &&  0.87 &&  1.00 &&  1.00 &&  0.96 &&  1.00 &&  1.00 &&  0.75 &&  1.00 &&  1.00 &&  0.95 &&  1.00 &&  1.00 &&  0.99 &&  1.00 &&  1.00 \\
200 &&  1.00 &&  1.00 &&  1.00 &&  0.95 &&  1.00 &&  1.00 &&  0.99 &&  1.00 &&  1.00 &&  0.90 &&  1.00 &&  1.00 &&  0.99 &&  1.00 &&  1.00 &&  1.00 &&  1.00 &&  1.00 \\
250 &&  1.00 &&  1.00 &&  1.00 &&  0.98 &&  1.00 &&  1.00 &&  1.00 &&  1.00 &&  1.00 &&  0.97 &&  1.00 &&  1.00 &&  1.00 &&  1.00 &&  1.00 &&  1.00 &&  1.00 &&  1.00 \\
300 &&  1.00 &&  1.00 &&  1.00 &&  1.00 &&  1.00 &&  1.00 &&  1.00 &&  1.00 &&  1.00 &&  0.99 &&  1.00 &&  1.00 &&  1.00 &&  1.00 &&  1.00 &&  1.00 &&  1.00 &&  1.00 \\
\multicolumn{37}{c}{}\\
\hline
\hline
\end{tabular}
\end{table}

\vspace{1in}

\setlength{\tabcolsep}{2pt}
\begin{table}[!htp]
\centering
\footnotesize
\caption{
        Power comparison at level 0.05 for the case in Section \ref{s:numerical} with $\{\tau, (\eta-\sigma, \eta+\sigma)\} =\{0.75, (0.25, 0.75)\}$,
        where $\mbox{CS}_1$ and $\mbox{CS}_2$ are the corresponding circularized versions defined in
        \eqref{eq:CS-mean-T} and \eqref{eq:CS-max-T}.
        }
        \label{tbl:pow-4}
\begin{tabular}{rcrcrcrcrcrcrcrcrcrcrcrcrcrcrcrcrcrcr}
\multicolumn{37}{c}{}\\
\hline\hline
\multicolumn{37}{c}{}\\
$n$&& \multicolumn{5}{c}{$W_n^2$}&& \multicolumn{5}{c}{$R_n^2$}
&& \multicolumn{5}{c}{AD}&& \multicolumn{5}{c}{LR}
&& \multicolumn{5}{c}{CvM}&& \multicolumn{5}{c}{KS}\\
\cline{1-1}
\cline{3-7}\cline{9-13}\cline{15-19}\cline{21-25}\cline{27-31}\cline{33-37}
&& && $\mbox{CS}_{1}$ && $\mbox{CS}_{2}$&& && $\mbox{CS}_{1}$ && $\mbox{CS}_{2}$&& && $\mbox{CS}_{1}$ && $\mbox{CS}_{2}$&&&& $\mbox{CS}_{1}$ && $\mbox{CS}_{2}$&&&& $\mbox{CS}_{1}$ && $\mbox{CS}_{2}$&& && $\mbox{CS}_{1}$ && $\mbox{CS}_{2}$\\
\cline{5-5}\cline{7-7}
\cline{11-11}\cline{13-13}
\cline{17-17}\cline{19-19}
\cline{23-23}\cline{25-25}
\cline{29-29}\cline{31-31}
\cline{35-35}\cline{37-37}\multicolumn{37}{c}{}\\
10 &&  0.08 &&  0.13 &&  0.13 &&  0.08 &&  0.14 &&  0.14 &&  0.08 &&  0.13 &&  0.13 &&  0.07 &&  0.14 &&  0.15 &&  0.10 &&  0.13 &&  0.14 &&  0.14 &&  0.14 &&  0.14 \\
50 &&  0.28 &&  0.71 &&  0.77 &&  0.17 &&  0.77 &&  0.79 &&  0.32 &&  0.72 &&  0.80 &&  0.16 &&  0.78 &&  0.81 &&  0.38 &&  0.70 &&  0.79 &&  0.60 &&  0.77 &&  0.77 \\
100 &&  0.64 &&  0.98 &&  0.99 &&  0.34 &&  0.99 &&  0.99 &&  0.63 &&  0.98 &&  0.99 &&  0.27 &&  0.99 &&  0.99 &&  0.72 &&  0.97 &&  0.99 &&  0.91 &&  0.99 &&  0.99 \\
150 &&  0.87 &&  1.00 &&  1.00 &&  0.54 &&  1.00 &&  1.00 &&  0.88 &&  1.00 &&  1.00 &&  0.45 &&  1.00 &&  1.00 &&  0.93 &&  1.00 &&  1.00 &&  0.99 &&  1.00 &&  1.00 \\
200 &&  0.97 &&  1.00 &&  1.00 &&  0.69 &&  1.00 &&  1.00 &&  0.96 &&  1.00 &&  1.00 &&  0.59 &&  1.00 &&  1.00 &&  0.99 &&  1.00 &&  1.00 &&  1.00 &&  1.00 &&  1.00 \\
250 &&  1.00 &&  1.00 &&  1.00 &&  0.83 &&  1.00 &&  1.00 &&  0.99 &&  1.00 &&  1.00 &&  0.75 &&  1.00 &&  1.00 &&  1.00 &&  1.00 &&  1.00 &&  1.00 &&  1.00 &&  1.00 \\
300 &&  1.00 &&  1.00 &&  1.00 &&  0.93 &&  1.00 &&  1.00 &&  1.00 &&  1.00 &&  1.00 &&  0.86 &&  1.00 &&  1.00 &&  1.00 &&  1.00 &&  1.00 &&  1.00 &&  1.00 &&  1.00 \\
\multicolumn{37}{c}{}\\
\hline
\hline
\end{tabular}
\end{table}





\end{document}